# Reflecting on Motivations: How Reasons to Publish affect Research Behaviour in Astronomy

**Julia Heuritsch**
Research Group “Reflexive Metrics”, Institut für Sozialwissenschaften, Humboldt Universität zu Berlin, Universitätsstraße 3B, 10117 Berlin, Germany; julia.heuritsch@hu-berlin.de, julia.heuritsch@gmail.com

## Abstract

Recent research in the field of reflexive metrics have studied the emergence and consequences of evaluation gaps in science. The concept of evaluation gaps captures potential discrepancies between what researchers value about their research, in particular research quality, and what metrics measure. As a result, scientists may experience anomie and adopt innovative ways to cope. These often value quantity over quality and may even compromise research integrity. A consequence of such gaps may therefore be research misconduct and a decrease in research quality. In the language of rational choice theory, an evaluation gap persists if motivational factors arising out of the internal component of an actor’s situation are incongruent with those arising out of the external components. The aim of this research is therefore to study and compare autonomous and controlled motivations to become an astronomer, to do research in astronomy and to publish scientific papers. Moreover, we study how these different motivational factors affect publication pressure, the experience of organisational justice and the observation of research misconduct. In summary, we find evidence for an evaluation gap and that controlled motivational factors arising from evaluation procedures based on publication record drives up publication pressure, which, in turn, was found to increase the likelihood of perceived frequency of misbehaviour.

## 1. Introduction

*Reflexive metrics* is a strand in science studies which explores how the demand for accountability and performance measurement in science has shaped the research culture in recent decades (e.g. Fochler & De Rijcke, 2017; Heuritsch, 2019). Drawing on organizational culture theories (OCT) and rational choice theory (RCT), we can analyse how the prevailing research culture in turn shapes research behaviour. Hypercompetition and publication pressure, which are shown to be part of the neoliberal research culture, may lead researchers to engage in scientific misconduct, compromising research integrity (e.g. Anderson et al, 2007; Halffman & Radder, 2015). Research integrity, in turn, has been linked to research quality (e.g. Martinson et al., 2010; Crain et al., 2013). While outright misconduct, such as fabrication, falsification or plagiarism (FFPs) is rather rare, it may be the many more subtle questionable research practices (QRPs), which jeopardize the quality of the knowledge produced (Martinson et al., 2005; Bouter et al., 2016; Haven; 2021). Concerns regarding research quality are raised and studied within the field of *reflexive metrics*.

Heuritsch (2021a) performed the first study bringing together the strand of research on the influence of the organisational culture on research integrity and *reflexive metrics*. By means of a quantitative survey, the author analysed how role-associated factors (such as academic position and location of employment) and cultural factors (such as perceived publication pressure and distributive & procedural justice) affect research behaviour and quality in astronomy. An astronomer who perceives publication pressure is more likely to perceive work less rewarding, less procedural justice and the necessity to put more effort into the work. Moreover, perceived publication pressure explains nearly 19% of the variance of observed misconduct. Among the forms of misconduct, QRPs are found to cause greater epistemic harm, because they occur more often than the more severe FFPs.

According to RCT, not only the prevailing culture, including its norms and rules, but also the present material opportunities, and personal motivations comprise an *actor's situation* (Esser, 1999). Personal motivations are part of the so-called internal component of an actor's situation, while culture, norms and material opportunities make up the three external components. The actor's situation influences their decisions and hence their behaviour.

A qualitative study by Heuritsch (2021b) found that astronomer's intrinsic motivation is to do research on the most fundamental questions of the universe out of curiosity and to "push knowledge forward". By contrast, their extrinsic motivation relates to complying with institutional norms, such as publishing a specific amount of papers per year, which is found to be the most important currency in the field. This may lead to a perceived evaluation gap; a discrepancy between what researchers value about their research, in particular research quality, and what metrics measure. The author found evidence that astronomers hence experience an anomie; they want to follow their intrinsic motivation to pursue science in order to push knowledge forward, while at the same time following their extrinsic motivation to comply with institutional norms. This may result in a balancing act to cope with the evaluation gap: the art of serving performance indicators in order to stay in academia, while at the same time compromising research quality as little as possible.

Building up on the qualitative research and quantitative research by Heuritsch (2021b & 2021a, respectively) we study different motivations for becoming an astronomer, for doing research in astronomy and for publishing scientific papers resulting from that research. Our aim is to compare these motivational factors and study how intrinsic and extrinsic motivations relate to each other. A discrepancy may indicate the presence of the evaluation gap and resulting balancing act found by Heuritsch (2021b). We therefore aim at completing the rational choice analysis of the structural conditions in astronomy, by paying tribute to the motivational factors arising out of the internal component of the astronomers' research situation. We quantitatively compare this autonomous motivation with the controlled motivation arising out of the external components. We are doing so by bringing together the strands of reflexive metrics, RCT, OCT and self-determination theory (SDT).

This paper will be structured as follows: First, we will give a theoretical background on SDT's conceptualization of motivation and an account for the link between motivation and behaviour. Second, the method section describes the sample selection, the survey instruments, research question & hypotheses and technical aspects concerning the statistical analyses. Third, the result section contains the results from our EFAs, CFAs, descriptive statistics, and our regression models. The result section is followed by a discussion, strength & limitations and a conclusion section that also gives an outlook for future studies.

## 2. Theoretical Background

### *2.1. The Motivation Continuum*

This study's operationalisation of the concept of motivation is rooted in self-determination theory (SDT; Deci & Ryan, 1985; Ryan & Connel, 1989: p.752, Gagné et al., 2010 & 2015), which belongs to the domain of organisational psychology. SDT proposes a multidimensional conceptualization of motivation and offers explanations of how these different types of motivation can be en- or discouraged (Gagné et al., 2015). At first order of approximation, SDT distinguishes between *extrinsic* and *intrinsic* motivation. Motivated by extrinsic motivation, one engages in an action for instrumental reasons, whereas an action performed out of intrinsic motivation is done for its own purpose. Extrinsic motivation is driven by values and goals, whereas intrinsic motivation is based on emotions, such as joy and curiosity.

At second order of approximation, SDT distinguishes between four different forms of extrinsic motivation. We may picture those four so-called *regulations* as a continuum (*Figure 1*), depending on their *degree of internalisation*. Internalisation refers to the assimilation of a regulation with existing self-regulations, based on one's values and interests. In other words, the higher the degree of internalisation of a regulation, the more does one identify with the value or meaning of the resulting activity. *External regulation*, the type of extrinsic motivation with the least degree of internalization, refers to doing an activity to obtain rewards, avoid punishments and includes rule-following. Next on the spectrum, we find *introjected regulation*, which refers to "the regulation of behaviour through self-worth contingencies such as ego-involvement and guilt" (Gagné et al., 2010: p.2). We act out of introjected regulation to receive (self-)approval and avoid disapproval. Introjected regulation is based on assimilating a regulation in a way that it becomes internally pressuring, which means it is partially internalised, but remains *controlling*. Identified regulation is almost fully internalised and hence one identifies with the value or meaning of the engaged activity. The goal that the activity poses is accepted as one's own and as such has personal importance. Therefore, identified regulation is not controlled, but *autonomous*. The most autonomous and completely internalised form of extrinsic motivation is *integrated regulation*. It "refers to identifying with the value of an activity to the point that it becomes part of a person's habitual functioning and part of the person's sense of self." (ibid.: p.2)

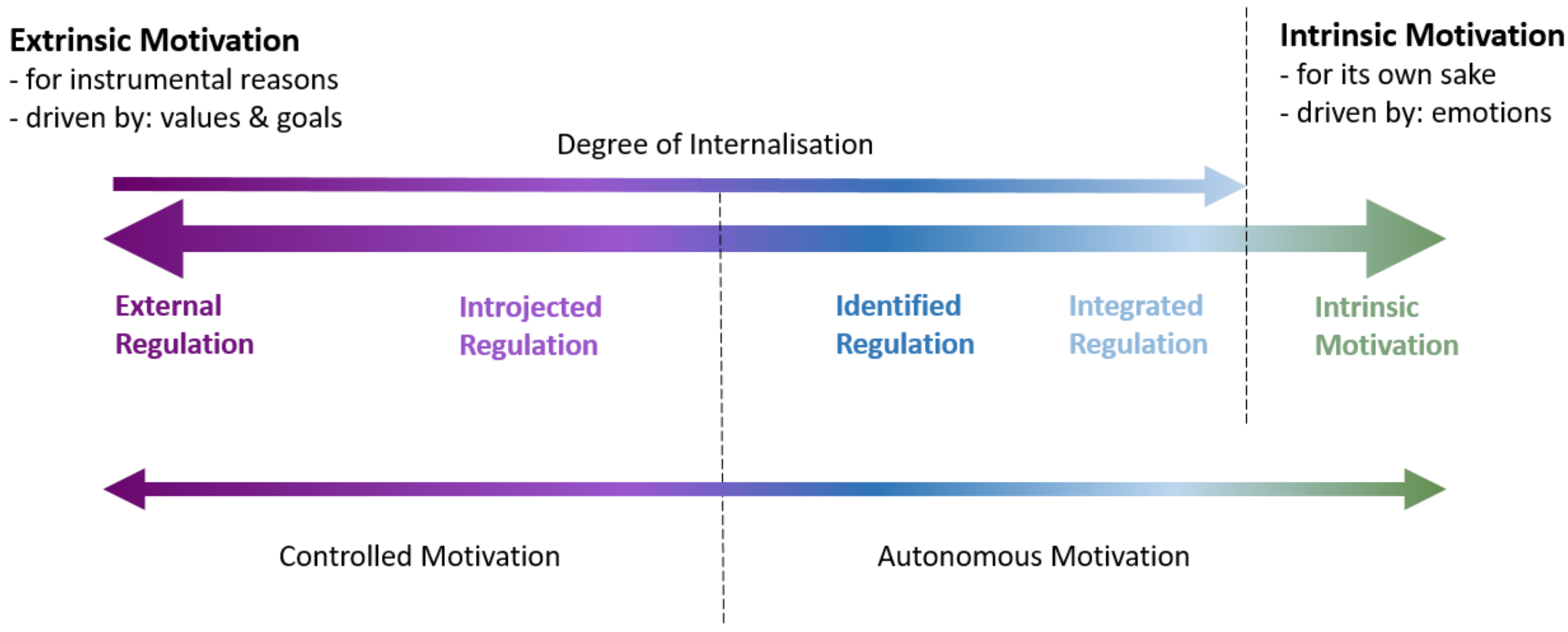

**Figure 1**. The motivation continuum (created by the author, based on Gagné et al., 2010 & 2015).

It is important to note that in practice one engages in an activity out of a mix of these five forms of motivation (four forms of extrinsic motivation and intrinsic motivation), with one of them at peak. In other words, when we say an action is motivated by, for example, introjected regulation, then this is the type of motivation that was most determining making the decision to engage in the action. Research has found that each subscale of motivation correlates most positively with adjacent subscales and less positively (or more negatively) with non-adjacent ones (Gagné et al., 2010).

The five different forms of motivation proposed by SDT make sense from a theoretical point of view, given the different degrees of internalisation of the regulations and that intrinsic motivation is the only form of motivation not based on instrumental reasons. However, in practice, we may not always be able to (statistically) differentiate within autonomous forms of motivation (identified regulation, integrated regulation and intrinsic motivation) and controlled forms of regulation (external and introjected). Research has shown that there is a clear break in the consequences of controlled versus autonomous types of motivation, hence for some research questions these aggregates may be sufficient (Gagné et al., 2010).

### *2.2. Reflexive Motivation: Impact on Behaviour*

In analogy to Heuritsch (2021b) we relate autonomous motivational factors to those that stem from the internal component (values & personal motives) of the actor's situation and controlled motivational factors to those arising out of the three external components (material opportunities, norms & culture). These motivational factors influence the decisions made by the actor and hence their behaviour, according to RCT (Esser, 1999). In other words, knowing which type of motivation is most present in a situation helps predicting the behavioural outcome. In turn, SDT-based organizational research (e.g. Gagné et al., 2015) shows that acting out of different types of motivation yields different forms of impact on the actor. According to Gagné et al., 2015, autonomous types of motivation are positively related to the satisfaction of psychological needs for autonomy. In SDT, "autonomy means endorsing one's actions at the highest level of reflection" (Gagné & Deci, 2005). This is one of three basic psychological needs for humans to function optimally, both in terms of their job-performance and their well-being. For example, autonomy is related to a higher job satisfaction, commitment, competence, personal initiative & involvement, effort, work motivation and less emotional distress, role stress, absenteeism and turnover intentions (Gagné et al., 2010).

The prevailing organisational climate is an important factor to foster or inhibit autonomous types of motivation. For example, motivational job design has the potential to compensate for poor leadership and booster motivation levels (Gagné et al., 2010). In other words, if employees feel in control of the situation (through job autonomy), managerial constraints do not necessarily have negative effects on their motivation. Participative management, which naturally increases perceived job autonomy, is found to have a positive impact on internalisation of the importance of the task, and subsequently leads to a greater intrinsic motivation & performance and less strain (ibid.).

By contrast, controlled motivation, unsurprisingly, is shown to be unrelated to autonomy support and need satisfaction, but related to other types of controlling leadership behaviours (Gagné et al., 2015). While the authors also found that introjected regulation was positively related to aspired outcomes, when controlling for identified motivation, many of these relations disappeared. Generally, acting out of controlled motivation is less beneficial to individuals' optimal functioning than autonomous motivation and may even lead to unwanted behaviour, such as deviant or unethical actions (ibid.).

## 3. Methods

### *3.1. Sample Selection and procedure*

This study is based on a web-based quantitative survey. The ideal aim was to run a census. However, because there is no official, complete list of all astronomers worldwide we used a multi-stage cluster sampling technique, targeting as many astronomers as possible. This consisted of distributing the survey invitation among astronomers from 176 universities, 56 non-academic research facilities & observatories and 17 societies & associations. In a consecutive stage, we asked the division heads of the International Astronomical Union (IAU) to distribute the survey invitation among their members. Five of nine division heads were non-responsive and so we reached the members of the remaining divisions through an automated script, based on publicly available email addresses. We estimate that around 13,000-15,000 astronomers were reached in total. 3509 astronomers completed the survey at least partly, amounting to a response rate of roughly 25%, and 2011 astronomers completed the survey in full. A full description of the sample selection and procedure is described elsewhere (Heuritsch, 2021a).

### *3.2. Instruments*

We used the online tool LimeSurvey to create and host the survey. As outlined above, the survey is embedded in the conceptual framework of rational choice theory (RCT), organisational culture theory (OCT) and self-determination theory (SDT). We used the same instruments as Heuritsch 2021a, but enhanced their model (*Figure 1 therein*) with motivational factors as independent variables. Therefore, this study is based on the following instruments to measure our independent and dependent variables:

#### 3.2.1. Dependent Variables

- <u>Scientific (Mis-) Behaviour</u>
  The instrument is described in detail elsewhere (Heuritsch, 2021a: p.7) and is originally based on Martinson et al. (2005, 2006, 2009, 2010) and Bouter et al. (2016). It measures the perceived frequency of observed (mis-) behaviour and consists of 18 items.

- <u>Turnover intentions & Love for job</u>
  In addition to instruments used in the analysis performed by Heuritsch (2021a), in this analysis we use three more dependent variables: Turnover intentions of academics & of non-academics and astronomer's love for their job. To measure these, we asked to what extent respondents agree/disagree to the following statements: "I regularly contemplate leaving academia" (for those primarily employed in academia), "I regularly contemplate quitting my job" (for those primarily employed in non-academic institutions) and "I love my job". The scale ranged from "Strongly Disagree" (1) to "Strongly Agree" (5).

#### 3.2.2. Independent Variables

- Perceived Publication Pressure
  The PPQ; the instrument measuring perceived publication pressure is described in detail elsewhere (Heuritsch, 2021a: p.7) and is originally based on Tijdink et al. (2014). The cleaned version of the construct we used here consists of 12 items.

- Perceived Organisational Justice: Distributive & Procedural Justice
  We measure perceived distributive justice via the Effort-Reward Imbalance (ERI; Siegrist et al., 2014) instrument. A detailed description of the instrument can be found elsewhere (Heuritsch, 2021a: p.7-8) and it consists of three effort and eight reward items. As for procedural justice (PJ), in this study we consider the following processes: a) Resource allocation, b) Peer review, c) Grant application and d) Telescope time application. The adaptation from the original instrument from Martinson et al. (2006) can be found elsewhere (Heuritsch, 2021a: p.7-8).

- Perceived Overcommitment
  The instrument measuring perceived overcommitment (OC), the inclination to overwork, is originally based on the ERI instrument (Siegrist et al., 2014) and the adaptation we used in this study can be found elsewhere (Heuritsch, 2021a: p.8).

- Motivation to become an Astronomer (M1):
  We designed 8 items for this instrument, which are inspired by Gagné et al. (2015). Five of these items intend to represent autonomous motivation and three items draw on controlled motivation, a classification we subsequently tested in a factor analysis. The full instrument can be found in *Table S1 in S1.*

- Motivation to publish:
  We used two instruments to measure this concept. In order to measure the "drivers to publish" (M2) we designed 9 items. Five of these items intend to represent autonomous motivation and four items draw on controlled motivation, a classification we subsequently tested in a factor analysis. The second instrument to measure the motivation publish does so by accounting for the feelings one experiences when not publishing the amount of papers that one aimed to publish (M3). This instrument consists of 10 items in total, where four items each intend to measure introjected regulation and external regulation. One item draws on identified regulation[1] and one item was used as a control item. Both instruments can be found in *Table S2a,b in S1* and are inspired by Gagné et al. (2015).

#### 3.2.3. Control Variables

Because we are enhancing Heuritsch's (2021a) model, we used the same role-associated and individual aspects as control variables: gender, academic position, whether one is primarily employed at an academic versus non-academic institution, whether one is employed at an institution in the global North/ South[2], and number of published first or co-author papers in the last 5 years. The reference categories are: gender: female/ non-binary; academic position: full professor; primary employer: Non-Academic; institute location: global South; Number of papers published in the last five years: 1-5.

---

[1] We note at this point that we did not include a subscale for integrated regulation in our study. This is because Gagné et al. (2015) explain that it can hardly be statistically separated from identified and intrinsic motivation subscales.

[2] https://meta.wikimedia.org/wiki/List_of_countries_by_regional_classification (accessed on 26th November 2021).

#### 3.2.4. Additional Instruments

In addition to our dependent, independent and control variables, that we include in our regression models, we are interested in two more aspects related to publication pressure and motivational aspects regarding work. We developed a nine-item battery to ask for the source of the perceived publication pressure (see *Table S3 in S1*) and asked respondents to rank the three most rewarding aspects of their work, given eight items (see *Table S4 in S1*).

### *3.3.Research Question & Hypotheses*

In the light of the previous research performed on the structural conditions of research in astronomy and theoretical background as outlined above, we work from the assumption that higher autonomous types of motivation the lower the likelihood to perceive publication pressure and the higher the likelihood to perceive organisational justice and to feel an overcommitment to work; and the other way round with controlled types of motivation. The relation between autonomous motivation and overcommitment is based on the fact that Martinson et al. (2006) conflate overcommitment with "intrinsic drive".

We have two research questions: 1) Is there a difference in the effect of autonomous versus controlled types of motivation on perceived publication pressure, distributive justice, overcommitment and frequency of observed misbehaviour? 2) Is the motivation for astronomers to perform research congruent with their motivation to publish papers?

Building up on the qualitative study on structural conditions of research in astronomy (Heuritsch, 2021b), the quantitative study on the relationship between perceived publication pressure, organisational justice, overcommitment and research misbehaviour in astronomy (Heuritsch, 2021a), as well as previous studies on organisational psychology and SDT (Gagné & Deci, 2005; Gagné et al. 2010 & 2015), this study tests the following hypotheses:

1. (H1): Astronomers take on that profession more out of autonomous than controlled motivation.
2. (H2): Astronomers' controlled motivation to publish is bigger than their autonomous motivation.
3. (H3): Those for whom publication demands pose a more serious threat to their academic career (e.g., early-career & female researchers in a male-dominated field) will feel less autonomous motivation and more controlled motivation to publish.
4. (H4): Higher autonomous motivation to become an astronomer and to publish decrease, while controlled motivation to become and astronomer and to publish increase the likelihood for astronomers to consider leaving academia/ quitting their job.
5. (H5): Higher autonomous motivation to become an astronomer and to publish increase, while controlled motivation to become and astronomer and to publish decrease the likelihood for astronomers to report that they love their job.
6. (H6): The perception of publication pressure increases with increasing controlled motivation and decreases with increasing autonomous motivation to publish.
7. (H7): The perception of effort put into work increases with increasing controlled motivation and decreases with increasing autonomous motivation to publish.

8. (H8): The perception of reward obtained from one's work increases with decreasing controlled motivation and with increasing autonomous motivation to publish.
9. (H9): The perception of overcommitment increases with increasing autonomous motivation to become an astronomer and with an increasing controlled motivation, external, introjected and identified regulation to publish.
10. (H10): The perception of frequency of misconduct decreases with an autonomous motivation to become an astronomer and to publish and increases with a controlled motivation to become an astronomer, and a controlled motivation, external and introjected regulation to publish.

### *3.4. Statistical Analyses*

The analysis of the survey data for this study was performed in SPSS and R. A full description of the data preparation can be found elsewhere (Heuritsch 2021a). All instruments, measuring the independent variables, are scored on a scale from 1 (strongly disagree) to 5 (strongly agree) and are treated as continuous variables in our regression models. The steps to arrive at our final model started with testing the independent variable constructs PPQ, ERI (including overcommitment) and M1-M3 by performing an exploratory factor analysis (EFA) for ordinal data (CATPCA in SPSS) and we derived Cronbach Alphas as scale reliabilities. For the EFA we used Promax Kaiser-normalisation for rotating the factors. Next, all independent variable constructs were tested by means of confirmatory factor analysis (CFA) using Lavaan version 0.5-23 (Rosseel, 2012) in R version 3.3.1. For each construct, we used the residual correlation matrices to determine significant correlations of the indicators and included them into the respective models. After checking for construct validity, we used a cleaned version of the PPQ, the effort-, reward- and overcommitment subscales and the factors of the motivational constructs as independent variables. Finally, we used SPSS linear regression analysis with listwise exclusion to test our hypotheses. The results section will present the results of our EFAs, CFAs, descriptive statistics, effects of the control variables on our motivational constructs and the regression models of turnover intentions & love for job, PPQ, ERI, OC and perceived misconduct being regressed onto the motivational factors and control variables.

## 4. Results

### *4.1. Exploratory & Confirmatory Factor Analyses*

The results of the EFAs & CFAs for the independent variables PPQ and ERI can be found elsewhere (Heuritsch, 2021a). This section describes the results of the EFAs & CFAs for the variables related to motivation: "Motivation to become an Astronomer" (M1), "Drivers to publish" (M2) and "Feelings when failing to publish" (M3).

As for M1, we set the CATPCA to yield four factors to account for the four different types of motivation (external, introjected & identified regulation and intrinsic motivation). The results can be found in *TableS1a in S2* and compared with the categorisation as expected from theory (see *TableS1b in S2*). An interesting outcome of this comparison is that the theoretical categorisation of external and introjected regulation matched the outcome of the CATPCA, while intrinsic motivation and identified regulation were switched around for four out of five items. This confirms the difficulty to distinguish between intrinsic motivation and identified regulation in practice. When testing the Cronbach alphas (see *TableS1c in S2*) for the whole construct we found that removing two items would increase the reliability of the M1 construct.

We subsequently decided to remove those items, resulting in 6 remaining items. The Cronbach alpha for the remaining M1 construct is 0.648 (see *TableS1d in S2*), which is not optimal, but workable. We henceforth used this cleaned M1 for any further analysis. We performed another CATPCA yielding two factors (see *TableS1e in S2*), which can be interpreted as M1F1: Autonomous Motivation to become an astronomer (comprising of identified regulation & intrinsic motivation) & M1F2: Controlled Motivation to become an astronomer (comprising of introjected regulation).

For M2 we also performed an exploratory 4-factor CATPCA (see *TableS2a in S2*) in order to compare our categorisation of the types of motivation based on theory with the results from the EFA. Only one item displays a difference and that is again an item that was theorised as belonging to identified regulation, but CATPCA associated it with intrinsic motivation (see *TableS2b in S2*). The Cronbach Alpha for the M2 construct yields 0.652 motivation (see *TableS2c in S2*), which in analogy to the M1 construct is workable. A 2-factor CATPCA (see *TableS2d in S2*) delivers M2F1: Autonomous Motivation to publish (comprising of identified regulation & intrinsic motivation) & M2F2: Controlled Motivation to publish (comprising of external & introjected regulation).

For M3 we performed an exploratory 3-factor CATPCA (see *TableS3a in S2*) in order to compare our categorisation of the types of motivation based on theory with the results from the EFA. The comparison (see *TableS3b in S2*) yields no difference. After removing the control item from the list, we performed an analysis of Cronbach's Alpha of the cleaned M3 construct, which amounts to 0.871 (see *TableS3c in S2*). M3 therefore shows good internal consistency. Performing another 3-factor CATPCA (see *TableS3d in S2*) for the cleaned M3 construct yields the following factors: M3F1: Identified Regulation to publish, M3F2: Introjected Regulation to publish & M3F3: External Regulation to publish.

We subsequently ran CFAs for all three of our motivation constructs M1-M3 and their factors, which results are presented in *TableS4 in S2*. This table includes the model fit indices CFI and TLI, where >0.9 indicates a good fit for both and RMSEA, where <0.05 denotes a good fit. All independent variable constructs show a good fit according to CFI and TLI. As for RMSEA the fit is good for M1 and acceptable for the other two constructs. Pearson correlation coefficients show weak to moderate correlations between the various motivational factors, but strong correlations between M2F2 and M3F2 (0.413), between M2F2 and M3F3 (0.5) and between M3F2 and M3F3 (0.503). This makes sense, since the controlled motivation to publish (M2F2) should be congruent with the introjected (M3F2) and external (M3F3) to publish. Furthermore, M3F2 and M3F3 are adjacent subscales of the motivation continuum and are therefore also expected to correlate positively (cf. Gagné et al. 2010).

### *4.2. Descriptive Statistics*

The descriptive statistics of the control variables (gender, academic position, primary employment at an academic/ non-academic institution, location of employment and number of published papers), independent variables (perceived publication pressure, perceived organisational distributive & procedural justice and perceived overcommitment), as well as the dependent variable perceived occurrence of misconduct can be found elsewhere (Heuritsch, 2021a).

*Table 1* presents the mean scores of the motivational factors, resulting from the factor analyses of our independent variable motivation constructs "Motivation to become an Astronomer" (M1), "Drivers to publish" (M2) and "Feelings when failing to publish" (M3). In addition to these independent variables, *Table 1* contains the mean scores of the dependent variables love for job and turnover intentions.

**Table 1**. Means and standard deviations of our independent variables – the factors of our motivation constructs "Motivation to become an Astronomer" (M1), "Drivers to Publish" (M2) & "Feelings when failing to Publish" (M3) – and dependent variables – turnover intentions & love for job. N = 3509 survey respondents and "n valid" excludes missing data.

| | n valid | Mean | SD |
|---|---|---|---|
| **Independent Variables** | | | |
| M1F1: Autonomous Motivation to become an astronomer | 2503 | 4.29 | 0.63 |
| M1F2: Controlled Motivation to become an astronomer | 2496 | 3.05 | 1.16 |
| M2F1: Autonomous Motivation to publish | 2032 | 3.50 | 0.70 |
| M2F2: Controlled Motivation to publish | 2032 | 3.76 | 0.77 |
| M3F1: Identified Motivation to publish | 1902 | 2.81 | 1.18 |
| M3F2: Introjected Motivation to publish | 1856 | 3.64 | 1.03 |
| M3F3: External Motivation to publish | 1919 | 3.19 | 1.17 |
| **Dependent Variables** | | | |
| Turnover Intentions (Academics) | 1467 | 2,53 | 1,45 |
| Turnover Intentions (Non-Academics) | 286 | 2,08 | 1,20 |
| Love for Job | 1759 | 4,13 | 0,92 |

*Table 1* shows that the autonomous motivation to become an astronomer is higher than the controlled one (means=4.29 versus 3.05, respectively) and the opposite is true for the motivation to publish (means=3.5 versus 3.76, respectively). The two forms of controlled motivation, introjected & external regulation to publish are bigger drivers that identified regulation (means=3.64, 3.19 & 2.81, respectively). The descriptive statistics of the items that make up factors M1F1 and M1F2 (see *Table S1 in S3*) yield that the comparatively biggest driver to become an astronomer is the "enjoyment of the process of gaining insight in astronomical phenomena" (mean=4.55), followed by "liking the intellectual challenge" (mean=4.48) and "finding out more about the laws that govern the universe" (mean=4.17), while the smallest driver is "needing a job" (mean=2.3). The descriptive statistics of the items that make up factors M2F1 and M2F2 (see *Table S2 in S3*) yield that the comparatively biggest driver to publish is an autonomous one, "publishing is important to share results with the community" (mean=4.51), closely followed by "publishing is a requirement from my job" (mean=3.97) and "publishing increases my reputation as a scientist" (mean=3.97). Similarly, respondents' comparatively biggest worries when not publishing the amount of papers that they aimed to publish (see *Table S3 in S3*) concern a possible negative impact on their career: "I am worried that it will decrease my chances for receiving external grants" (mean=3.91), "I am worried that it will negatively impact my career prospects" (mean=3.63) and "I am worried that it will negatively impact my research track record" (mean=3.63).

2500 astronomers responded to the ranking question "What do you find most rewarding about your work?". Most astronomers chose "enjoying the process of finding truths about the universe" (n=1076) for rank 1. N=652 chose "making incremental steps in building up knowledge" as rank 2 and n=702 selected "getting a paper published" as rank 3. When summing up all responses for all three ranks, we obtain the ranking presented in *Table 2*, were "making ground-breaking steps in building up knowledge" took the 3$^{rd}$ place before "getting a paper published". The least important rewards, both, in the individual rank answers and when summed up were chosen as "receiving a job promotion", "winning scientific prizes" and "receiving a salary raise".

**Table 2**. Ranking of what astronomers find most rewarding about their work.

| Rank | Item | n |
|---|---|---|
| 1 | Enjoying the process of finding truths about the universe | 1934 |
| 2 | Making incremental steps in building up knowledge | 1582 |
| 3 | Making ground-breaking steps in building up knowledge | 1398 |
| 4 | Getting a paper published | 1268 |
| 5 | Receiving praise from a colleague/ my supervisor | 609 |
| 6 | Receiving a job promotion (a more senior job title) | 316 |
| 7 | Receiving a salary raise | 207 |
| 8 | Winning scientific prizes | 182 |

1951 astronomers responded to the multiple choice question what the source of their perceived publication pressure is. Results are visualised in *Figure 2*. The two most chosen answers are "I need to maintain credibility as a scientist" (N=1167) and "I need to boost my publication record for increasing my career chance" (N=1128). The least chosen source of publication pressure were "I need to earn prestige" (N=412) and "I need to avoid failure" (N=436).

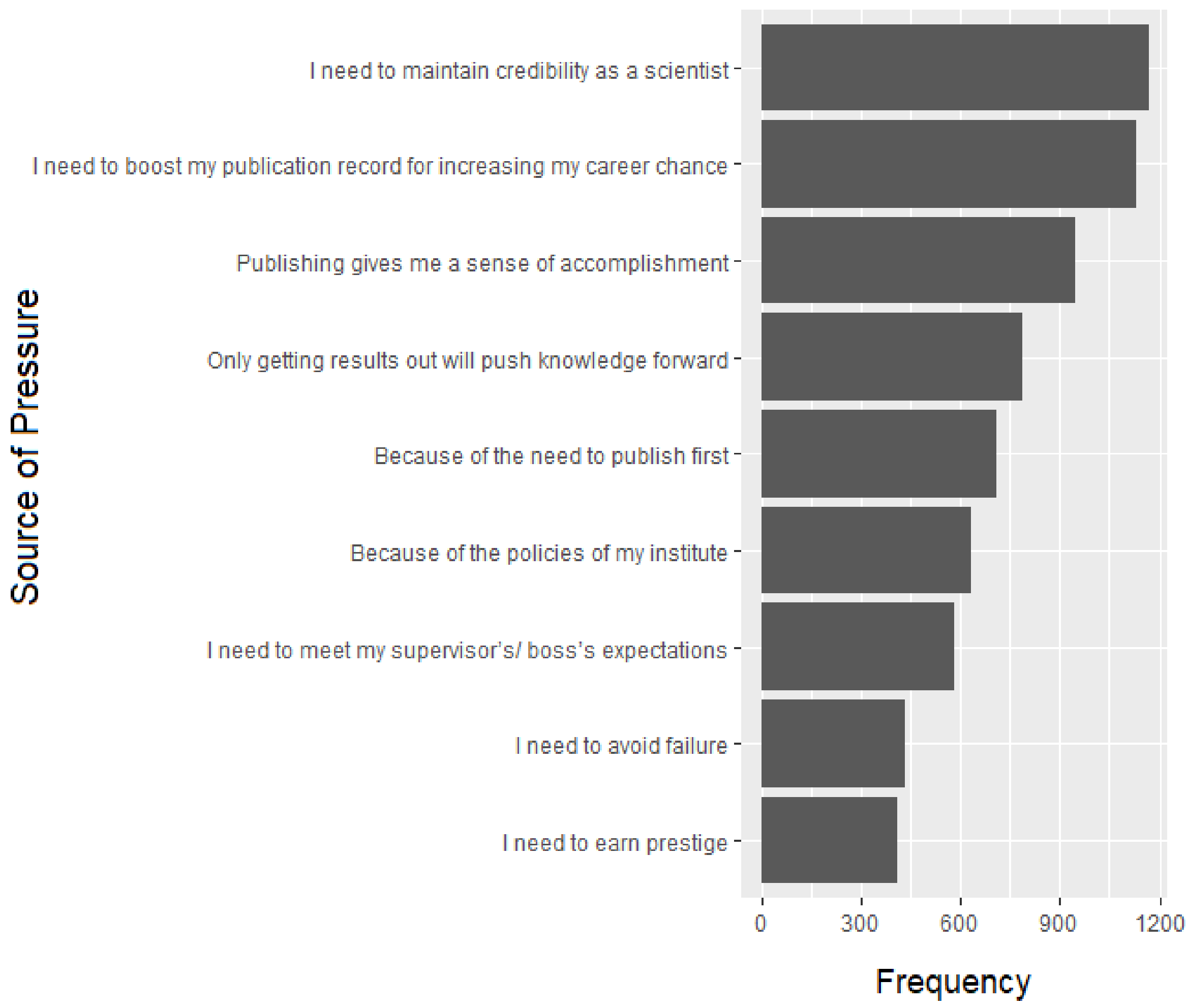

**Figure 2**. Source of perceived publication pressure, ranked by frequency of the respondents' answers.

*4.3. Motivations & Control Variables*

This section presents the results of the regression models that include the seven motivational factors (M1F1, M1F2, M2F1, M2F2, M3F1, M3F2 & M3F3) as dependent variables and the control variables as independent ones. As a reminder; reference categories for the control variables are: gender: female/ non-binary; academic position: full professor; primary employer: Non-Academic; institute location: Global South; number of papers published: 1–5).

*Table 3* (N=1360) presents the regression model for the autonomous motivation to become an astronomer in dependence of the control variables. Male as opposed to female/ non-binary tend to feel less autonomous motivation to become an astronomer (by 0.084 points). Having published more than 20 papers in the last 5 years is related to a higher autonomous motivation to become an astronomer (by 0.121 points).

**Table 3**. Regression model of DV = Autonomous Motivation to become an astronomer; regressed onto the control variables. * indicates statistical significance ($p < 0.05$).

| | | **Unstandardized Coefficients** | | **Standardized Coefficients** | **t** | **Sig.** |
|---|---|---|---|---|---|---|
| | | **B** | **Std. Error** | **Beta** | | |
| **M1F1: Autonomous Motivation to become an astronomer** | Intercept: | 4.287 | 0.098 | | 43.914 | <0.001 * |
| | Gender: Male | -0.084 | 0.039 | -0.061 | -2.183 | 0.029 * |
| | Position: PhD Candidate | -0.098 | 0.069 | -0.053 | -1.415 | 0.157 |
| | Position: Postdoc | -0.090 | 0.050 | -0.061 | -1.799 | 0.072 |
| | Position: Assistant Prof. | -0.089 | 0.068 | -0.039 | -1.306 | 0.192 |
| | Position: Associate Prof. | -0.041 | 0.056 | -0.023 | -0.731 | 0.465 |
| | Position: Other | -0.105 | 0.059 | -0.057 | -1.779 | 0.076 |
| | Primary Employer: Academic | 0.059 | 0.074 | 0.023 | 0.800 | 0.424 |
| | Papers published: Submission | -0.039 | 0.122 | -0.010 | -0.319 | 0.750 |
| | Papers published: 0 | -0.075 | 0.098 | -0.022 | -0.763 | 0.446 |
| | Papers published: 6–20 | 0.047 | 0.046 | 0.035 | 1.025 | 0.305 |
| | Papers published: >20 | 0.121 | 0.047 | 0.090 | 2.552 | 0.011 * |
| | Location: Global North | 0.021 | 0.047 | 0.012 | 0.449 | 0.653 |

The controlled motivation to become an astronomer (*Table 4*, N=1357) by contrast, tends to be higher for males as compared to females/ non-binaries (by 0.157 points). Postdocs and astronomers from the position category “Other” perceive less controlled motivation to become an astronomer than Full Professors (by 0.243 & 0.255 points, respectively). Astronomers employed by an institution in the Global North also report less controlled motivation to become an astronomer as compared to the Global South (by 0.203 points).

**Table 4**. Regression model of DV = Controlled Motivation to become an astronomer; regressed onto the control variables. * indicates statistical significance ($p < 0.05$).

| | | **Unstandardized Coefficients** | | **Standardized Coefficients** | **t** | **Sig.** |
|---|---|---|---|---|---|---|
| | | **B** | **Std. Error** | **Beta** | | |
| **M1F2: Controlled Motivation** | Intercept: | 3.083 | 0.178 | | 17.343 | <0.001 * |
| | Gender: Male | 0.157 | 0.070 | 0.062 | 2.239 | 0.025 * |
| | Position: PhD Candidate | -0.092 | 0.126 | -0.028 | -0.736 | 0.462 |
| **to become an astronomer** | Position: Postdoc | -0.243 | 0.092 | -0.090 | -2.657 | 0.008 * |
| | Position: Assistant Prof. | -0.191 | 0.124 | -0.046 | -1.545 | 0.123 |
| | Position: Associate Prof. | -.085 | 0.102 | -0.026 | -0.833 | 0.405 |
| | Position: Other | -0.255 | 0.108 | -0.076 | -2.367 | 0.018 * |
| | Primary Employer: Academic | 0.237 | 0.134 | 0.050 | 1.770 | 0.077 |
| | Papers published: Submission | -.094 | 0.222 | -0.013 | -0.422 | 0.673 |
| | Papers published: 0 | 0.116 | 0.178 | 0.019 | 0.653 | 0.514 |
| | Papers published: 6–20 | -0.124 | 0.083 | -0.050 | -1.496 | 0.135 |
| | Papers published: >20 | -0.080 | 0.086 | -0.033 | -.925 | 0.355 |
| | Location: Global North | -0.203 | 0.085 | -0.065 | -2.382 | 0.017 * |

The autonomous motivation to publish (*Table 5*, N=1361) tends to be lower for males as compared to females/ non-binaries (by 0.095 points). Positions ranked lower than Full Professor and "Other" are also likely to feel less autonomous motivation to publish as compared to Full Professors, except from Associate Professors, whose effect is not statistically significant. Astronomers who have published more than 20 papers in the last 5 years feel more autonomous motivation to publish than those who have published 1 to 5 papers. Those employed by an institution in the Global North report less autonomous motivation to publish than those on the Global South (by 0.328 points).

**Table 5**. Regression model of DV = Autonomous Motivation to publish; regressed onto the control variables. * indicates statistical significance ($p < 0.05$).

| | | **Unstandardized Coefficients** | | **Standardized Coefficients** | **t** | **Sig.** |
|---|---|---|---|---|---|---|
| | | **B** | **Std. Error** | **Beta** | | |
| **M2F1: Autonomous Motivation to publish** | Intercept: | 3.984 | 0.104 | | 38.269 | <0.001 * |
| | Gender: Male | -0.095 | 0.041 | -0.062 | -2.309 | 0.021 * |
| | Position: PhD Candidate | -0.268 | 0.074 | -0.132 | -3.640 | <0.001 * |
| | Position: Postdoc | -0.259 | 0.054 | -0.158 | -4.830 | <0.001 * |
| | Position: Assistant Prof. | -0.160 | 0.072 | -0.064 | -2.209 | 0.027 * |
| | Position: Associate Prof. | -0.109 | 0.060 | -0.055 | -1.827 | 0.068 |
| | Position: Other | -0.199 | 0.063 | -0.098 | -3.149 | 0.002 * |
| | Primary Employer: Academic | -0.046 | 0.079 | -0.016 | -0.586 | 0.558 |
| | Papers published: Submission | -0.092 | 0.130 | -0.020 | -0.705 | 0.481 |
| | Papers published: 0 | 0.090 | 0.105 | 0.025 | 0.862 | 0.389 |
| | Papers published: 6–20 | 0.063 | 0.049 | 0.042 | 1.296 | 0.195 |
| | Papers published: >20 | 0.159 | 0.051 | 0.108 | 3.149 | 0.002 * |
| | Location: Global North | -0.328 | 0.050 | -0.176 | -6.589 | <0.001 * |

The controlled motivation to publish (*Table 6*, N=1361) also tends to be lower for males than females/ non-binaries (by 0.225 points). Contrary to the autonomous motivation to publish, positions ranked lower than Full Professor tend to perceive more controlled motivation to publish than Full Professors. Astronomers who haven't published any papers in the last 5 years feel less controlled motivation to publish than those having published 1 to 5 papers (by 0.257 points).

**Table 6**. Regression model of DV = Controlled Motivation to publish; regressed onto the control variables. * indicates statistical significance ($p < 0.05$).

| | | Unstandardized Coefficients | | Standardized Coefficients | t | Sig. |
|---|---|---|---|---|---|---|
| | | B | Std. Error | Beta | | |
| **M2F2: Controlled Motivation to publish** | Intercept: | 3.756 | 0.113 | | 33.261 | <0.001 * |
| | Gender: Male | -0.225 | 0.045 | -0.135 | -5.034 | <0.001 * |
| | Position: PhD Candidate | 0.405 | 0.080 | 0.183 | 5.077 | <0.001 * |
| | Position: Postdoc | 0.386 | 0.058 | 0.216 | 6.648 | <0.001 * |
| | Position: Assistant Prof. | 0.261 | 0.078 | 0.096 | 3.331 | 0.001 * |
| | Position: Associate Prof. | 0.147 | 0.065 | 0.068 | 2.282 | 0.023 * |
| | Position: Other | -0.071 | 0.068 | -0.032 | -1.043 | 0.297 |
| | Primary Employer: Academic | 0.076 | 0.085 | 0.024 | 0.895 | 0.371 |
| | Papers published: Submission | -0.046 | 0.141 | -0.009 | -0.328 | 0.743 |
| | Papers published: 0 | -0.257 | 0.113 | -0.064 | -2.268 | 0.023 * |
| | Papers published: 6–20 | 0.037 | 0.053 | 0.023 | 0.703 | 0.482 |
| | Papers published: >20 | 0.038 | 0.055 | 0.023 | 0.685 | 0.494 |
| | Location: Global North | -0.087 | 0.054 | -0.042 | -1.602 | 0.109 |

Identified regulation to publish (*Table 7*, N=1305) tends to be lower for males than females/ non-binaries (by 0.217 points) as well as for PhD Candidates as compared to Full Professors (by 0.307 points).

**Table 7**. Regression model of DV = Identified Regulation to publish; regressed onto the control variables. * indicates statistical significance ($p < 0.05$).

| | | Unstandardized Coefficients | | Standardized Coefficients | t | Sig. |
|---|---|---|---|---|---|---|
| | | B | Std. Error | Beta | | |
| **M3F1: Identified Regulation to publish** | Intercept: | 3.515 | 0.182 | | 19.268 | <0.001 * |
| | Gender: Male | -0.217 | 0.072 | -0.085 | -2.996 | 0.003 * |
| | Position: PhD Candidate | -0.307 | 0.130 | -0.090 | -2.368 | 0.018 * |
| | Position: Postdoc | -0.100 | 0.094 | -0.037 | -1.059 | 0.290 |
| | Position: Assistant Prof. | -0.013 | 0.127 | -0.003 | -0.099 | 0.921 |
| | Position: Associate Prof. | -0.048 | 0.105 | -0.015 | -0.454 | 0.650 |
| | Position: Other | -0.182 | 0.111 | -0.054 | -1.640 | 0.101 |
| | Primary Employer: Academic | -0.135 | 0.138 | -0.028 | -0.983 | 0.326 |
| | Papers published: Submission | 0.143 | 0.228 | 0.019 | 0.625 | 0.532 |
| | Papers published: 0 | 0.301 | 0.188 | 0.047 | 1.600 | 0.110 |
| | Papers published: 6–20 | 0.014 | 0.085 | 0.006 | 0.170 | 0.865 |
| | Papers published: >20 | -0.084 | 0.089 | -0.034 | -0.940 | 0.347 |
| | Location: Global North | 0.094 | 0.087 | 0.030 | 1.077 | 0.282 |

Introjected regulation to publish (*Table 8*, N=1294) is likely to be lower for males than females/non-binaries (by 0.416 points). PhD Candidates, Postdocs and Assistant Professors tend to have a higher introjected regulation to publish than Full Professors. Astronomers who currently have their first paper in the submission process feel a higher introjected motivation to publish than those who have published 1-5 papers in the last five years (by 0.538 points) and those who have published more than 20 papers feel a lower introjected motivation (by 0.310 points).

**Table 8**. Regression model of DV = Introjected Regulation to publish; regressed onto the control variables. * indicates statistical significance ($p < 0.05$).

| | | Unstandardized Coefficients | | Standardized Coefficients | t | Sig. |
|---|---|---|---|---|---|---|
| | | **B** | **Std. Error** | **Beta** | | |
| **M3F2: Introjected Regulation to publish** | Intercept: | 2.713 | 0.178 | | 15.276 | <0.001 * |
| | Gender: Male | -0.416 | 0.070 | -0.161 | -5.950 | <0.001 * |
| | Position: PhD Candidate | 0.450 | 0.125 | 0.131 | 3.609 | <0.001 * |
| | Position: Postdoc | 0.496 | 0.091 | 0.181 | 5.461 | <0.001 * |
| | Position: Assistant Prof. | 0.390 | 0.122 | 0.094 | 3.199 | 0.001 * |
| | Position: Associate Prof. | 0.170 | 0.102 | 0.051 | 1.672 | 0.095 |
| | Position: Other | 0.162 | 0.108 | 0.047 | 1.492 | 0.136 |
| | Primary Employer: Academic | 0.244 | 0.134 | 0.050 | 1.829 | 0.068 |
| | Papers published: Submission | 0.538 | 0.219 | 0.071 | 2.453 | 0.014 * |
| | Papers published: 0 | -0.043 | 0.179 | -0.007 | -0.242 | 0.809 |
| | Papers published: 6–20 | -0.105 | 0.082 | -0.042 | -1.281 | 0.200 |
| | Papers published: >20 | -0.310 | 0.086 | -0.123 | -3.602 | <0.001 * |
| | Location: Global North | 0.041 | 0.085 | 0.013 | 0.484 | 0.629 |

External regulation to publish (*Table 9*, N=1265) also tends to be lower for males than females/ non-binaries (by 0.399 points) and higher for all positions ranked lower than Full Professors. As with introjected motivation to publish, astronomers whose first paper is currently in the submission process feel a higher external regulation to publish (by 0.466 points) and those who have published more than 20 papers in the last five years feel a lower external regulation than the reference category of 1 to 5 papers (by 0.176 points). Astronomers being employed in the Global North tend to experience a higher external motivation to publish (by 0.213 points) than those in the Global South.

**Table 9**. Regression model of DV = External Regulation to publish; regressed onto the control variables. * indicates statistical significance ($p < 0.05$).

| | | Unstandardized Coefficients | | Standardized Coefficients | t | Sig. |
|---|---|---|---|---|---|---|
| | | **B** | **Std. Error** | **Beta** | | |
| **M3F3: External Regulation to publish** | Intercept: | 3.277 | 0.152 | | 21.618 | <0.001 * |
| | Gender: Male | -0.399 | 0.060 | -0.176 | -6.679 | <0.001 * |
| | Position: PhD Candidate | 0.506 | 0.108 | 0.164 | 4.677 | <0.001 * |
| | Position: Postdoc | 0.774 | 0.077 | 0.323 | 10.046 | <0.001 * |
| | Position: Assistant Prof. | 0.772 | 0.104 | 0.211 | 7.447 | <0.001 * |
| | Position: Associate Prof. | 0.404 | 0.086 | 0.140 | 4.705 | <0.001 * |
| | Position: Other | 0.165 | 0.093 | 0.054 | 1.773 | 0.077 |
| | Primary Employer: Academic | 0.163 | 0.115 | 0.038 | 1.412 | 0.158 |
| | Papers published: Submission | 0.466 | 0.195 | 0.068 | 2.386 | 0.017 * |
| | Papers published: 0 | -0.051 | 0.171 | -0.008 | -0.300 | 0.764 |
| | Papers published: 6–20 | -0.072 | 0.070 | -0.033 | -1.027 | 0.305 |
| | Papers published: >20 | -0.176 | 0.073 | -0.080 | -2.413 | 0.016 * |
| | Location: Global North | 0.213 | 0.072 | 0.077 | 2.967 | 0.003 * |

*4.4. Motivations & Turnover intentions and Love for Job*

Our regression analysis on which of our seven motivational factors play a role in regularly contemplating leaving academia yield the following results (*Table 10*; N=1147): Those astronomers who feel more autonomous motivation to become an astronomer and more autonomous motivation to publish feel significantly less often that they want to leave academia (by 0.283 & 0.239 points, respectively). In contrast, those who feel more introjected and external regulation to publish are more likely to consider moving on from academia (by 0.176 & 0.291 points, respectively). Perhaps unsurprisingly, astronomers from other positions than Full Professor have a much higher chance of leaving academia. Astronomers who haven't published a paper in the last five years, or those who have published more than 20 papers, tend to have less turnover intentions than those of the reference category (1-5 papers).

**Table 10**. Regression model of DV = Turnover intentions of astronomers who are primarily employed at an academic institution; regressed onto the seven motivational factors and the control variables. * indicates statistical significance ($p < 0.05$).

| | | **Unstandardized Coefficients** | | **Standardized Coefficients** | **t** | **Sig.** |
|---|---|---|---|---|---|---|
| | | **B** | **Std. Error** | **Beta** | | |
| **Turnover Intentions (Academics)** | Intercept: | 2.485 | 0.355 | | 6.990 | <0.001 * |
| | M1F1: Aut. mot. astronomer | -0.283 | 0.061 | -0.121 | -4.666 | <0.001 * |
| | M1F2: Contr. mot. astronomer | -0.026 | 0.033 | -0.020 | -0.789 | 0.430 |
| | M2F1: Aut. mot. publishing | -0.239 | 0.058 | -0.113 | -4.101 | <0.001 * |
| | M2F2: Contr. mot. publishing | 0.082 | 0.058 | 0.042 | 1.417 | 0.157 |
| | M3F1: Identified reg. publishing | -0.040 | 0.033 | -0.032 | -1.231 | 0.218 |
| | M3F2: Introjected reg. publishing | 0.176 | 0.037 | 0.144 | 4.804 | <0.001 * |
| | M3F3: External reg. publishing | 0.291 | 0.045 | 0.207 | 6.472 | <0.001 * |
| | Gender: Male | 0.068 | 0.083 | 0.021 | 0.829 | 0.408 |
| | Position: PhD Candidate | 0.873 | 0.148 | 0.202 | 5.912 | <0.001 * |
| | Position: Postdoc | 0.996 | 0.107 | 0.300 | 9.269 | <0.001 * |
| | Position: Assistant Prof. | 0.294 | 0.140 | 0.058 | 2.100 | 0.036 * |
| | Position: Associate Prof. | 0.231 | 0.115 | 0.057 | 2.005 | 0.045 * |
| | Position: Other | 0.322 | 0.130 | 0.068 | 2.475 | 0.013 * |
| | Papers published: Submission | 0.480 | 0.258 | 0.050 | 1.860 | 0.063 |
| | Papers published: 0 | -0.547 | 0.231 | -0.061 | -2.370 | 0.018 * |
| | Papers published: 6–20 | -0.097 | 0.093 | -0.032 | -1.047 | 0.295 |
| | Papers published: >20 | -0.197 | 0.099 | -0.063 | -1.985 | 0.047 * |
| | Location: Global North | 0.016 | 0.098 | 0.004 | 0.163 | 0.871 |

Our regression analysis on which of our seven motivational factors play a role in regularly contemplating quitting their job at their non-academic primary employer (*Table 11*; N=224) yield no statistically significant results for the motivational factors. However, we find that astronomers who have published more than 20 papers in the last five years are less likely (by 0.572 points) to think about a turnover as compared to those of the reference category (1-5 papers).

**Table 11**. Regression model of DV = Turnover intentions of astronomers who are primarily employed at a non-academic institution; regressed onto the seven motivational factors and the control variables. * indicates statistical significance ($p < 0.05$).

| | | **Unstandardized Coefficients** | | **Standardized Coefficients** | **t** | **Sig.** |
|---|---|---|---|---|---|---|
| | | **B** | **Std. Error** | **Beta** | | |
| **Turnover Intentions (Non-Academics)** | Intercept: | 2.394 | 0.804 | | 2.976 | 0.003 * |
| | M1F1: Aut. mot. astronomer | -0.040 | 0.133 | -0.021 | -0.304 | 0.762 |
| | M1F2: Contr. mot. astronomer | -0.023 | 0.075 | -0.021 | -0.307 | 0.760 |
| | M2F1: Aut. mot. publishing | -0.109 | 0.137 | -0.058 | -0.794 | 0.428 |
| | M2F2: Contr. mot. publishing | 0.027 | 0.152 | 0.017 | 0.180 | 0.857 |
| | M3F1: Identified reg. publishing | -0.012 | 0.079 | -0.012 | -0.153 | 0.879 |
| | M3F2: Introjected reg. publishing | 0.079 | 0.091 | 0.074 | 0.863 | 0.389 |
| | M3F3: External reg. publishing | 0.162 | 0.102 | 0.145 | 1.593 | 0.113 |
| | Gender: Male | -0.237 | 0.177 | -0.092 | -1.337 | 0.183 |
| | Papers published: Submission | -0.671 | 0.841 | -0.053 | -0.798 | 0.426 |
| | Papers published: 0 | -0.790 | 0.437 | -0.122 | -1.805 | 0.072 |
| | Papers published: 6–20 | -0.325 | 0.197 | -0.126 | -1.646 | 0.101 |
| | Papers published: >20 | -0.572 | 0.202 | -0.222 | -2.834 | 0.005 * |
| | Location: Global North | -0.036 | 0.308 | -0.008 | -0.118 | 0.906 |

Our regression analysis on which of our seven motivational factors play a role in loving their job (whether academic or non-academic) yield the following results (*Table 12*; N=1226): Those who feel more autonomous motivation to become an astronomer and to publish are more likely to love their job (by 0.268 & 0.259 points, respectively). The same goes for those who report more controlled motivation to become an astronomer and more identified regulation to publish (by 0.049 & 0.059 points, respectively). The opposite is true for astronomers who have a higher introjected or external regulation to publish (by 0.139 & 0.086 points, respectively) – these tend to love their jobs less. An astronomer whose paper is currently in the submission process also tends to love their job less (by 0.493 points).

**Table 12**. Regression model of DV = Love for Job; regressed onto the seven motivational factors and the control variables. * indicates statistical significance ($p < 0.05$).

| | | **Unstandardized Coefficients** | | **Standardized Coefficients** | **t** | **Sig.** |
|---|---|---|---|---|---|---|
| | | **B** | **Std. Error** | **Beta** | | |
| **Love for Job** | Intercept: | 2.690 | 0.262 | | 10.261 | <0.001 * |
| | M1F1: Aut. mot. astronomer | 0.268 | 0.041 | 0.178 | 6.520 | <0.001 * |
| | M1F2: Contr. mot. astronomer | 0.049 | 0.023 | 0.059 | 2.169 | 0.03 * |
| | M2F1: Aut. mot. publishing | 0.259 | 0.040 | 0.189 | 6.482 | <0.001 * |
| | M2F2: Contr. mot. publishing | 0.001 | 0.040 | 0.001 | 0.030 | 0.976 |
| | M3F1: Identified reg. publishing | 0.059 | 0.023 | 0.073 | 2.614 | 0.009 * |
| | M3F2: Introjected reg. publishing | -0.139 | 0.025 | -0.174 | -5.487 | <0.001 * |
| | M3F3: External reg. publishing | -0.086 | 0.031 | -0.094 | -2.786 | 0.005 * |
| | Gender: Male | 0.004 | 0.057 | 0.002 | 0.063 | 0.949 |
| | Position: PhD Candidate | -0.132 | 0.101 | -0.047 | -1.312 | 0.190 |
| | Position: Postdoc | 0.023 | 0.074 | 0.011 | 0.308 | 0.758 |
| | Position: Assistant Prof. | 0.102 | 0.097 | 0.031 | 1.047 | 0.295 |

| | | B | Std. Error | Beta | t | Sig. |
|---|---|---|---|---|---|---|
| | Position: Associate Prof. | 0.019 | 0.080 | 0.007 | 0.243 | 0.808 |
| | Position: Other | 0.066 | 0.086 | 0.024 | 0.772 | 0.440 |
| | Primary Employer: Academic | -0.186 | 0.105 | -0.047 | -1.768 | 0.077 |
| | Papers published: Submission | -0.493 | 0.176 | -0.080 | -2.792 | 0.005 * |
| | Papers published: 0 | 0.269 | 0.156 | 0.047 | 1.728 | 0.084 |
| | Papers published: 6–20 | 0.013 | 0.064 | 0.006 | 0.198 | 0.843 |
| | Papers published: >20 | -0.001 | 0.068 | -0.001 | -0.017 | 0.986 |
| | Location: Global North | -0.120 | 0.068 | -0.048 | -1.782 | 0.075 |

### *4.5. Motivations & PPQ, ERI, OC*

We enhanced the regression models of the analysis performed by Heuritsch (2021a) with our seven motivational factors as independent variables and present the results in this section.

The perception of publication pressure (*Table 13*; N=1212) significantly increases with increasing autonomous motivation to become an astronomer (by 0.098 points) and introjected & external regulation to publish (by 0.116 & 0.292 points, respectively). By contrast, astronomers who feel more autonomous motivation and identified regulation to publish tend to perceive less publication pressure (by 0.107 & 0.049 points, respectively). When directly comparing this regression model with the model by Heuritsch (2021a; *Table 3 therein*; N=520), we notice that the gender and current position of the astronomer have lost its statistical significance. This may be partly due the difference in N and partly due to the mediated effects of the added motivational factors. As evident from *section 4.3.*, lower ranked positions than Full Professor tend to feel a lower autonomous motivation to publish and a higher controlled motivation, introjected and external regulation to publish.

**Table 13**. Regression model of DV = Perceived Publication Pressure; regressed onto the seven motivational factors and the control variables. * indicates statistical significance ($p < 0.05$).

| | | **Unstandardized Coefficients** | | **Standardized Coefficients** | **t** | **Sig.** |
|---|---|---|---|---|---|---|
| | | **B** | **Std. Error** | **Beta** | | |
| **Perceived Publication Pressure (PPQ)** | Intercept: | 2.266 | 0.199 | | 11.396 | <0.001 * |
| | M1F1: Aut. mot. astronomer | 0.098 | 0.031 | 0.080 | 3.140 | 0.002 * |
| | M1F2: Contr. mot. astronomer | -0.009 | 0.017 | -0.013 | -0.526 | 0.599 |
| | M2F1: Aut. mot. publishing | -0.107 | 0.030 | -0.096 | -3.509 | <0.001 * |
| | M2F2: Contr. mot. publishing | -0.028 | 0.030 | -0.027 | -0.933 | 0.351 |
| | M3F1: Identified reg. publishing | -0.049 | 0.017 | -0.075 | -2.887 | 0.004 * |
| | M3F2: Introjected reg. publishing | 0.116 | 0.019 | 0.177 | 6.034 | <0.001 * |
| | M3F3: External reg. publishing | 0.292 | 0.023 | 0.391 | 12.501 | <0.001 * |
| | Gender: Male | -0.029 | 0.043 | -0.017 | -0.685 | 0.494 |
| | Position: PhD Candidate | 0.108 | 0.077 | 0.046 | 1.396 | 0.163 |
| | Position: Postdoc | 0.084 | 0.056 | 0.047 | 1.497 | 0.135 |
| | Position: Assistant Prof. | 0.057 | 0.073 | 0.021 | 0.779 | 0.436 |
| | Position: Associate Prof. | -0.061 | 0.060 | -0.028 | -1.007 | 0.314 |
| | Position: Other | 0.029 | 0.065 | 0.013 | 0.447 | 0.655 |
| | Primary Employer: Academic | 0.142 | 0.079 | 0.045 | 1.800 | 0.072 |
| | Papers published: Submission | -0.067 | 0.135 | -0.013 | -0.495 | 0.621 |
| | Papers published: 0 | -0.054 | 0.129 | -0.010 | -0.420 | 0.675 |
| | Papers published: 6–20 | -0.053 | 0.048 | -0.033 | -1.099 | 0.272 |
| | Papers published: >20 | -0.134 | 0.051 | -0.082 | -2.617 | 0.009 * |
| | Location: Global North | -0.373 | 0.052 | -0.180 | -7.214 | <0.001 * |

As for the distributive justice factors reward and effort (*Table 14*; N=1207 & N=1208, respectively), the perception of obtaining rewards from one's job increases with a higher controlled motivation to become an astronomer (by 0.041 points) and a higher autonomous & controlled motivation to publish (by 0.071 & 0.070 points, respectively). By contrast, respondents perceive their work less rewarding when they show a higher introjected and external regulation to publish (by 0.062 & 0.103 points, respectively). The perception of effort put into one's job significantly increases when respondents feel a higher controlled motivation, introjected & external regulation to publish (by 0.104, 0.089 & 0.132 points, respectively). By contrast, one tends to feel less need to put effort when one perceives a higher autonomous motivation to publish (by 0.135 points). The trend observed in the model by Heuritsch (2021a) that Full Professors tend to feel better rewarded than an astronomer from any other position is also observed here, although with larger effects. At the same time, ECRs tend to feel putting less effort into their work. Interestingly, astronomers, who have not published any papers in the last five years, feel obtaining more reward from and putting less effort into their jobs than those having published 1-5 papers in the same time frame.

**Table 14**. Regression models of DV = Distributive Justice in terms of Reward & Effort; regressed onto the seven motivational factors, perceived publication pressure, and the control variables. * indicates statistical significance ($p < 0.05$).

| | | **Unstandardized Coefficients** | | **Standardized Coefficients** | **t** | **Sig.** |
|---|---|---|---|---|---|---|
| | | **B** | **Std. Error** | **Beta** | | |
| **Distributive Justice (DJ): Reward** | Intercept: | 4.256 | 0.239 | | 17.840 | <0.001 * |
| | M1F1: Aut. mot. astronomer | 0.009 | 0.036 | 0.006 | 0.241 | 0.810 |
| | M1F2: Contr. mot. astronomer | 0.041 | 0.019 | 0.056 | 2.126 | 0.034 * |
| | M2F1: Aut. mot. publishing | 0.071 | 0.035 | 0.057 | 2.031 | 0.042 * |
| | M2F2: Contr. mot. publishing | 0.070 | 0.034 | 0.061 | 2.063 | 0.039 * |
| | M3F1: Identified reg. publishing | 0.020 | 0.020 | 0.027 | 1.018 | 0.309 |
| | M3F2: Introjected reg. publishing | -0.062 | 0.022 | -0.085 | -2.791 | 0.005 * |
| | M3F3: External reg. publishing | -0.103 | 0.028 | -0.124 | -3.645 | <0.001 * |
| | PPQ: Perceived Publication Pressure | -0.325 | 0.033 | -0.292 | -9.831 | <0.001 * |
| | Gender: Male | -0.044 | 0.049 | -0.024 | -0.909 | 0.363 |
| | Position: PhD Candidate | -0.451 | 0.088 | -0.173 | -5.109 | <0.001 * |
| | Position: Postdoc | -0.456 | 0.064 | -0.233 | -7.135 | <0.001 * |
| | Position: Assistant Prof. | -0.271 | 0.084 | -0.090 | -3.242 | 0.001 * |
| | Position: Associate Prof. | -0.223 | 0.069 | -0.093 | -3.255 | 0.001 * |
| | Position: Other | -0.176 | 0.074 | -0.070 | -2.391 | 0.017 * |
| | Primary Employer: Academic | -0.037 | 0.090 | -0.010 | -0.408 | 0.683 |
| | Papers published: Submission | 0.059 | 0.154 | 0.010 | 0.381 | 0.703 |
| | Papers published: 0 | 0.301 | 0.147 | 0.052 | 2.047 | 0.041 * |
| | Papers published: 6–20 | 0.034 | 0.055 | 0.019 | 0.610 | 0.542 |
| | Papers published: >20 | 0.108 | 0.058 | 0.060 | 1.847 | 0.065 |
| | Location: Global North | 0.040 | 0.060 | 0.017 | 0.662 | 0.508 |
| | Intercept: | 1.914 | 0.234 | | 8.182 | <0.001 * |
| | M1F1: Aut. mot. astronomer | 0.059 | 0.035 | 0.045 | 1.676 | 0.094 |
| **Distributive Justice (DJ): Effort** | M1F2: Contr. mot. astronomer | -0.034 | 0.019 | -0.047 | -1.753 | 0.080 |
| | M2F1: Aut. mot. publishing | -0.135 | 0.034 | -0.113 | -3.930 | <0.001 * |
| | M2F2: Contr. mot. publishing | 0.104 | 0.033 | 0.095 | 3.106 | 0.002 * |
| | M3F1: Identified reg. publishing | 0.032 | 0.019 | 0.045 | 1.658 | 0.098 |
| | M3F2: Introjected reg. publishing | 0.089 | 0.022 | 0.127 | 4.078 | <0.001 * |
| | M3F3: External reg. publishing | 0.132 | 0.028 | 0.165 | 4.726 | <0.001 * |
| | PPQ: Perceived Publication Pressure | 0.293 | 0.032 | 0.275 | 9.047 | <0.001 * |
| | Gender: Male | -0.016 | 0.048 | -0.009 | -0.329 | 0.743 |
| | Position: PhD Candidate | -0.394 | 0.087 | -0.157 | -4.543 | <0.001 * |
| | Position: Postdoc | -0.387 | 0.063 | -0.205 | -6.165 | <0.001 * |
| | Position: Assistant Prof. | -0.033 | 0.082 | -0.012 | -0.407 | 0.684 |
| | Position: Associate Prof. | -0.030 | 0.067 | -0.013 | -0.445 | 0.657 |
| | Position: Other | -0.172 | 0.072 | -0.071 | -2.384 | 0.017 * |
| | Primary Employer: Academic | -0.011 | 0.088 | -0.003 | -0.129 | 0.897 |
| | Papers published: Submission | -0.379 | 0.151 | -0.070 | -2.515 | 0.012 * |
| | Papers published: 0 | 0.025 | 0.144 | 0.004 | 0.170 | 0.865 |
| | Papers published: 6–20 | -0.061 | 0.054 | -0.035 | -1.124 | 0.261 |
| | Papers published: >20 | 0.096 | 0.057 | 0.055 | 1.666 | 0.096 |
| | Location: Global North | 0.048 | 0.059 | 0.022 | 0.807 | 0.420 |

Enhancing the regression model for perceived overcommitment with the seven motivational factors yields the results presented in *Table 15* (Model 1, N=591): Respondents with a higher introjected and identified regulation to publish feel more overcommitment (by 0.110 & 0.078 points; respectively). Taking perceived publication pressure and the perceived organizational justice items out of the equation (*Table 15*; Model 2, N=1226), we obtain significant effects for all motivational factors, apart from controlled motivation to become an astronomer and controlled motivation to publish. Perceived overcommitment then increases when respondents report a higher autonomous motivation to become an astronomer (by 0.116 points), and a higher introjected, external and identified regulation to publish (by 0.171, 0.2 and 0.049 points, respectively). Autonomous motivation to publish decreases the likelihood of perceived overcommitment (0.097 points).

**Table 15**. Regression models of DV = Overcommitment; Model 1: DV is regressed onto the seven motivational factors, perceived publication pressure, distributive justice, and the control variables. Model 2: DV is regressed onto the seven motivational factors and the control variables. * indicates statistical significance ($p < 0.05$).

| | | **Unstandardized Coefficients** | | **Standardized Coefficients** | **t** | **Sig.** |
|---|---|---|---|---|---|---|
| | | **B** | **Std. Error** | **Beta** | | |
| **Over-commitment (OC) Model 1** | Intercept: | 1.293 | 0.429 | | 3.016 | 0.003 * |
| | M1F1: Aut. mot. astronomer | 0.057 | 0.051 | 0.040 | 1.113 | 0.266 |
| | M1F2: Contr. mot. astronomer | 0.004 | 0.028 | 0.005 | 0.145 | 0.885 |
| | M2F1: Aut. mot. publishing | -0.068 | 0.053 | -0.050 | -1.284 | 0.200 |
| | M2F2: Contr. mot. publishing | -0.018 | 0.047 | -0.015 | -0.370 | 0.711 |
| | M3F1: Identified reg. publishing | 0.078 | 0.027 | 0.105 | 2.873 | 0.004 * |
| | M3F2: Introjected reg. publishing | 0.110 | 0.030 | 0.152 | 3.671 | <0.001 * |
| | M3F3: External reg. publishing | 0.024 | 0.041 | 0.027 | 0.583 | 0.560 |
| | PPQ: Perceived Publication Pressure | 0.199 | 0.050 | 0.177 | 3.961 | <0.001 * |
| | DJ: Effort | 0.396 | 0.043 | 0.350 | 9.185 | <0.001 * |
| | DJ: Reward | -0.059 | 0.047 | -0.059 | -1.251 | 0.211 |
| | OJ: Resource Allocation | -0.031 | 0.045 | -0.030 | -0.681 | 0.496 |
| | OJ: Peer Review | 0.042 | 0.047 | 0.036 | 0.903 | 0.367 |
| | OJ: Grant Application | -0.012 | 0.044 | -0.011 | -0.261 | 0.794 |
| | OJ: Telescope Application | -0.071 | 0.048 | -0.059 | -1.476 | 0.140 |
| | Gender: Male | -0.020 | 0.068 | -0.011 | -0.294 | 0.769 |
| | Position: PhD Candidate | 0.166 | 0.168 | 0.037 | 0.988 | 0.324 |
| | Position: Postdoc | -0.034 | 0.091 | -0.017 | -0.377 | 0.707 |
| | Position: Assistant Prof. | -0.050 | 0.104 | -0.018 | -0.483 | 0.629 |
| | Position: Associate Prof. | -0.195 | 0.088 | -0.086 | -2.223 | 0.027 * |
| | Position: Other | -0.109 | 0.100 | -0.043 | -1.087 | 0.278 |
| | Primary Employer: Academic | -0.054 | 0.115 | -0.016 | -0.467 | 0.641 |
| | Papers published: 6–20 | 0.070 | 0.082 | 0.038 | 0.851 | 0.395 |
| | Papers published: >20 | 0.082 | 0.081 | 0.048 | 1.019 | 0.309 |
| | Location: Global North | 0.024 | 0.088 | 0.010 | 0.272 | 0.785 |
| **Over-commitment (OC)** **Model 2** | Intercept: | 2.131 | 0.250 | | 8.522 | <0.001 * |
| | M1F1: Aut. mot. astronomer | 0.116 | 0.039 | 0.082 | 2.946 | 0.003 * |
| | M1F2: Contr. mot. astronomer | -0.034 | 0.022 | -0.043 | -1.558 | 0.120 |
| | M2F1: Aut. mot. publishing | -0.097 | 0.038 | -0.076 | -2.554 | 0.011 * |
| | M2F2: Contr. mot. publishing | 0.008 | 0.038 | 0.007 | 0.205 | 0.838 |
| | M3F1: Identified reg. publishing | 0.049 | 0.021 | 0.065 | 2.293 | 0.022 * |
| | M3F2: Introjected reg. publishing | 0.171 | 0.024 | 0.228 | 7.091 | <0.001 * |
| | M3F3: External reg. publishing | 0.200 | 0.029 | 0.233 | 6.789 | <0.001 * |
| | Gender: Male | -0.021 | 0.054 | -0.011 | -0.388 | 0.698 |
| | Position: PhD Candidate | -0.094 | 0.096 | -0.036 | -0.982 | 0.327 |
| | Position: Postdoc | -0.154 | 0.071 | -0.076 | -2.184 | 0.029 * |
| | Position: Assistant Prof. | -0.158 | 0.093 | -0.051 | -1.707 | 0.088 |
| | Position: Associate Prof. | -0.179 | 0.076 | -0.072 | -2.354 | 0.019 * |
| | Position: Other | -0.183 | 0.082 | -0.069 | -2.238 | 0.025 * |
| | Primary Employer: Academic | 0.072 | 0.100 | 0.020 | 0.718 | 0.473 |
| | Papers published: Submission | -0.324 | 0.168 | -0.056 | -1.926 | 0.054 |
| | Papers published: 0 | -0.188 | 0.149 | -0.035 | -1.266 | 0.206 |
| | Papers published: 6–20 | 0.065 | 0.061 | 0.034 | 1.059 | 0.290 |
| | Papers published: >20 | 0.177 | 0.065 | 0.094 | 2.745 | 0.006 * |
| | Location: Global North | -0.148 | 0.064 | -0.062 | -2.294 | 0.022 * |

*4.6. Motivations & Scientific Misbehaviour*

In a final step, we regressed our seven motivational factors and the control variables on frequency of misbehavior occurrence (*Table 16*; Model 2, N=1208). We observe that controlled motivation to become an astronomer and autonomous motivation to publish decrease the observation of misbehaviour (by 0.038 & 0.067 points; respectively). By contrast, introjected & external regulation to publish increase the perceived frequency of misbehaviour occurrence (by 0.085 & 0.144 points; respectively). When adding the PPQ, ERI, OC and PJ constructs to the regression model (*Table 16*; Model 1, N=586,), as done by Heuritsch (2021a), the effects of the motivational factors yield no statistical significance anymore, apart from the controlled motivation to become an astronomer (decrease of 0.045 points), emphasising the mediating effect that the cultural constructs have (*section 4.5.*). In contrast to the analysis performed by Heuritsch (2021a), increased perceived procedural justice in terms of peer review also decreases the likelihood to observe misbehaviour (0.093 points).

**Table 16**. Regression models of DV = Perception of Frequency of Misbehaviour Occurence; Model 1: DV is regressed onto the seven motivational factors, perceived publication pressure, distributive justice, overcommitment, and the control variables. Model 2: DV is regressed onto the seven motivational factors and the control variables. * indicates statistical significance ($p < 0.05$).

| | | **Unstandardized Coefficients** | | **Standardized Coefficients** | **t** | **Sig.** |
|---|---|---|---|---|---|---|
| | | **B** | **Std. Error** | **Beta** | | |
| **Perception of misbehaviour occurrence Model 1** | Intercept: | 2.911 | 0.347 | | 8.393 | <0.001 * |
| | M1F1: Aut. mot. astronomer | -0.018 | 0.041 | -0.016 | -0.435 | 0.664 |
| | M1F2: Contr. mot. astronomer | -0.045 | 0.022 | -0.079 | -2.012 | 0.045 * |
| | M2F1: Aut. mot. publishing | 0.045 | 0.043 | 0.043 | 1.056 | 0.292 |
| | M2F2: Contr. mot. publishing | -0.003 | 0.038 | -0.003 | -0.074 | 0.941 |
| | M3F1: Identified reg. publishing | 0.021 | 0.022 | 0.037 | 0.955 | 0.34 |
| | M3F2: Introjected reg. publishing | 0.044 | 0.024 | 0.079 | 1.8 | 0.072 |
| | M3F3: External reg. publishing | 0.007 | 0.033 | 0.01 | 0.21 | 0.834 |
| | PPQ: Perceived Publication Pressure | 0.262 | 0.041 | 0.304 | 6.45 | <0.001 * |
| | DJ: Effort | 0.07 | 0.037 | 0.08 | 1.873 | 0.062 |
| | DJ: Reward | -0.071 | 0.038 | -0.093 | -1.884 | 0.06 |
| | OJ: Resource Allocation | 0.009 | 0.036 | 0.011 | 0.253 | 0.8 |
| | OJ: Peer Review | -0.093 | 0.038 | -0.104 | -2.474 | 0.014 * |
| | OJ: Grant Application | -0.031 | 0.036 | -0.037 | -0.879 | 0.38 |
| | OJ: Telescope Application | -0.145 | 0.038 | -0.159 | -3.77 | <0.001 * |
| | OC: Overcommitment | 0.004 | 0.034 | 0.005 | 0.107 | 0.915 |
| | Gender: Male | 0.004 | 0.054 | 0.003 | 0.071 | 0.944 |
| | Position: PhD Candidate | 0.147 | 0.135 | 0.043 | 1.092 | 0.275 |
| | Position: Postdoc | 0.079 | 0.073 | 0.051 | 1.084 | 0.279 |
| | Position: Assistant Prof. | 0.162 | 0.083 | 0.078 | 1.959 | 0.051 |
| | Position: Associate Prof. | 0.012 | 0.071 | 0.007 | 0.164 | 0.87 |
| | Position: Other | -0.012 | 0.08 | -0.006 | -0.15 | 0.881 |
| | Primary Employer: Academic | -0.15 | 0.092 | -0.06 | -1.627 | 0.104 |
| | Papers published: 6–20 | -0.013 | 0.067 | -0.01 | -0.203 | 0.84 |
| | Papers published: >20 | 0.041 | 0.065 | 0.031 | 0.626 | 0.531 |
| | Location: Global North | 0.166 | 0.07 | 0.089 | 2.367 | 0.018 * |
| **Perception of misbehaviour occurrence Model 2** | Intercept: | 2.759 | 0.188 | | 14.671 | <0.001 * |
| | M1F1: Aut. mot. astronomer | 0.04 | 0.03 | 0.039 | 1.364 | 0.173 |
| | M1F2: Contr. mot. astronomer | -0.038 | 0.016 | -0.065 | -2.312 | 0.021 * |
| | M2F1: Aut. mot. publishing | -0.067 | 0.029 | -0.07 | -2.336 | 0.02 * |
| | M2F2: Contr. mot. publishing | -0.037 | 0.028 | -0.042 | -1.285 | 0.199 |
| | M3F1: Identified reg. publishing | -0.019 | 0.016 | -0.033 | -1.153 | 0.249 |
| | M3F2: Introjected reg. publishing | 0.085 | 0.018 | 0.153 | 4.687 | <0.001 * |
| | M3F3: External reg. publishing | 0.144 | 0.022 | 0.226 | 6.481 | <0.001 * |
| | Gender: Male | -0.068 | 0.041 | -0.047 | -1.673 | 0.095 |
| | Position: PhD Candidate | 0.159 | 0.073 | 0.081 | 2.188 | 0.029 * |
| | Position: Postdoc | 0.14 | 0.053 | 0.092 | 2.62 | 0.009 * |
| | Position: Assistant Prof. | 0.256 | 0.07 | 0.11 | 3.652 | <0.001 * |

| | | | | | |
|---|---|---|---|---|---|
| | Position: Associate Prof. | -0.044 | 0.057 | -0.024 | -0.762 | 0.446 |
| | Position: Other | -0.029 | 0.062 | -0.015 | -0.474 | 0.636 |
| | Primary Employer: Academic | -0.114 | 0.075 | -0.042 | -1.514 | 0.13 |
| | Papers published: Submission | 0.049 | 0.128 | 0.011 | 0.383 | 0.702 |
| | Papers published: 0 | -0.079 | 0.118 | -0.019 | -0.672 | 0.501 |
| | Papers published: 6–20 | -0.011 | 0.046 | -0.008 | -0.242 | 0.809 |
| | Papers published: >20 | 0.052 | 0.049 | 0.037 | 1.064 | 0.288 |
| | Location: Global North | -0.022 | 0.049 | -0.013 | -0.462 | 0.644 |

## 5. Discussion

This study's quantitative results supports findings of the qualitative research performed by Heuritsch (2021b) and we find evidence supporting our hypotheses. The author (ibid.) found that astronomers' biggest reasons to take on their profession are their intrinsic motivation to follow their curiosity about the universe and the intellectual challenge of studying astrophysical phenomena. Indeed, we find that curiosity and enjoyment of the intellectual challenge are crucial motivational factors for our respondents to become an astronomer and the comparatively biggest one is the "enjoyment of the process of gaining insight in astronomical phenomena". Our findings indicate that autonomous motivation to become an astronomer dominates the controlled one (H1).

Consistent with our findings on motivational factors to become an astronomer, our respondents ranked as most rewarding about their work to be "enjoying the process of finding truths about the universe" and "making incremental & ground-breaking steps in building up knowledge". Moreover, the comparatively biggest driver to publish is the importance "to share results with the community". These results are congruent with astronomer's autonomous motivation to push knowledge forward (cf. Heuritsch, 2021b); Astronomers enjoy (intrinsic motivation) to unravel the mysteries of the universe and find it important to share their results to push knowledge forward, while not enjoying "the process writing a paper" and the paper review process as much (comparatively smallest driver to publish). This means that their comparatively biggest driver to publish is one out of identified regulation – to get the results to the community for others to build on the knowledge. As hypothesised (H2), however, controlled autonomous dominates over autonomous motivation to publish. These controlled motivational factors mainly relate to the need of papers for one's career. Hence, while respondents rank "getting a paper published" only as a medium rewarding experience, our results show that it has significant importance for their career.

The importance of publishing for astronomers' career is emphasised by the fact that most respondents reported the need to boost their publication record for increasing their career chances as the biggest source for publication pressure. By contrast, a third less respondents reported that "only getting results out will push knowledge forward" sparks their publication pressure, even though this source of pressure would fit more with their autonomous motivation to push knowledge forward.

The comparison between motivational factors of becoming an astronomer and to publish leads us to conclude a disagreement in values: While the autonomous motivation to become an astronomer is bigger than the controlled one, the controlled motivation to publish is bigger than the autonomous one. We therefore find quantitative support for the evaluation gap and anomie found by Heuritsch (2021b) – a discrepancy between the importance of autonomous motivation to push knowledge forward and the controlled motivation to produce the most valued outcomes (publications).

When analysing what role-associated and individual aspects play a role in the extent of the seven motivational factors we deducted from our factor analysis, we find that men feel a lower autonomous motivation and a higher controlled motivation for becoming an astronomer than females/ non-binaries. We can only speculate about the reasons: In a field that is still male-dominated, women may feel more that they have chosen astrophysics by their own accord, out of curiosity, as opposed to due to external factors. By contrast, the controlled motivation, as well as external, introjected and identified regulations to publish are found to be higher for females/ non-binaries in comparison to males. This supports our hypothesis (H3), however only partly, since the autonomous motivation to publish also tends to be higher for females/ non-binaries than for males. The trend we find when looking at how the academic position of an astronomer influences the seven motivational factors also supports hypothesis (H3): higher academic positions are generally related to a higher autonomous motivation to publish. By contrast, the controlled motivation as well as the introjected and external regulations to publish are found to be lower for astronomers with a higher academic rank. This is consistent with the finding of Heuritsch (2021a) that those with a higher rank perceive less publication pressure compared to those with a lower rank. It makes sense that astronomers who feel more controlled than autonomous motivation to perform a certain action also feel more pressure to do so. Relatedly, astronomers who haven't published at all feel a higher introjected and external regulation to publish as compared to those who have published 1-5 papers in the last five years. The opposite is true for those who have published more than 20 papers in the last five years. Lastly, we found that astronomers employed by an institution in the Global North report less controlled motivation to become an astronomer and less autonomous motivation to publish. External regulation to publish, by contrast, is found higher as compared to those primarily employed in the Global South. Interestingly however, Heuritsch (2021a) found that astronomers employed in the Global North feel less publication pressure than those employed in the Global South. The reasons for the differences of Global North versus Global South deserve further investigation.

Gagné et al. (2010) report on the negative relation between the feeling of work autonomy and turnover intentions, which inspired hypothesis H4. We found indeed that astronomers who report a higher autonomous motivation to become an astronomer and a higher autonomous motivation to publish contemplate less often to leave academia. In contrast, controlled motivation, as well as external and introjected regulation to publish increases the likelihood for an astronomer to consider leaving academia. As for primarily non-academic astronomers the results for the influence of motivational factors on turnover were not statistically significant. We also found support for our hypothesis H5; a higher autonomous motivation to become an astronomer and a higher autonomous & identified motivation to publish increase the likelihood of astronomers to report that they love their job and the opposite is true for a higher introjected and external regulation to publish. Additionally, a higher controlled motivation to become an astronomer also increases the chance that one is happy with their job.

The perception of publication pressure increases with introjected and external regulation to publish and decreases with autonomous motivation and identified regulation to publish, which supports hypothesis H6. The results for the perception of effort put into one's work are comparable (hypothesis H7): perceived effort increases when one feels a higher controlled motivation and introjected & external regulation to publish and decreases with a higher autonomous motivation to publish. The results for the perception of obtaining rewards from one's job are inconclusive: while a higher autonomous and controlled motivation to publish increase the likelihood to feel rewarded, a higher introjected and external regulation to publish lead astronomers to perceive their work less rewarding. We cannot explain this discrepancy between controlled motivation versus introjected and external regulation to publish, since the former by definition comprises the latter two concepts and – as we have shown – these concepts correlate positively, as expected. Therefore, hypothesis H8 is not supported. As for perceived overcommitment, our hypothesis H9 is partly supported: taking into account that publication pressure, perceived effort & reward and procedural justice act as a mediator, a higher introjected and identified regulation to publish have direct positive effects on overcommitment. As for perception of frequency of misbehaviour, we find H10 partly supported: while we do find evidence for autonomous motivation to publish to decrease the perception of misconduct occurrence, and introjected and external regulation to publish to lead to an increase, we find that controlled motivation to become an astronomer decreases, instead of increases, this perception. When accounting for PPQ, ERI, OC and OJs as mediators (such as in the model by Heuritsch, 2021a), the only direct effect from a motivational factor that is statistically significant is controlled motivation to become an astronomer.

## 6. Strengths/ Limitations

A detailed analysis of the strengths and limitations of the web-based survey this study is based on is described elsewhere (Heuritsch, 2021a). For this study specifically, we point out that the Cronbach Alphas for the motivation constructs could be improved by enhancing the item batteries. Especially for M3 there is only one item drawing on identified regulation. Despite extensive search in literature, we found it challenging to design more items representing this kind of motivation for the question "How do you feel when you don't publish the amount of papers that you aimed to publish?".

Furthermore, we observe slight differences in the regression results compared to the study by Heuritsch (2021a). These may stem from the fact, that for the analysis in this study we used SPSS's linear regression algorithm, instead of specifying and calculating our whole SEM model with Lavaan in R (leading to differences in N). We chose this simpler method as adding the additional seven motivational constructs into our SEM would make the model too complex and we would risk overspecification.

## 7. Conclusion, Implications and Outlook for Further Research

The aim of this research was to study the motivations of people becoming an astronomer and to publish as an output of their work. We distinguish between autonomous, arising of the internal component of an actor's situation, and controlled types of motivation, arising of the external components of the actor's situation – the culture, norms and material opportunities present at action. We found a discrepancy between the importance of publications as a personal reward and their importance for astronomer's career. In other words, while astronomers' biggest autonomous goal is to push knowledge forward (cf. Heuritsch, 2021b), their biggest drivers to publish are related to their career prospects. Hence, what performance indicators, such as the publication rate, measure diverges from what astronomers value. This evaluation gap could indeed play into the balancing act found by Heuritsch (2021b), where astronomers aim at producing quality science that pushes knowledge forward, while at the same time sacrificing just as much quality that is necessary to also produce the expected quantity of papers.

Recent research (e.g. Gagné & Deci, 2005; Gagné et al., 2010 & 2015) on organisational psychology shows that autonomous motivation is related to better work outcomes (e.g. performance, well-being, competence etc) and indeed we find that a higher autonomous motivation to publish decreases perception of publication pressure while a higher introjected & external regulation to publish increase this perceived pressure. The same goes for perceived effort put into work. As Heuritsch (2021a) shows, perceived publication pressure, in turn, increases the perception of a higher scientific misconduct. We conclude that driving up factors of controlled motivation through incentives based on performance indicators, such as publication rate, may have negative effects on astronomer's well-being and research conduct. Moreover, while identified regulation to publish increases overcommitment, it decreases publication pressure and increases how much astronomers like their job. As Gagné et al. (2015) points out, studies have shown that performance was more highly correlated with identified than with intrinsic motivation. Given this and our findings, we suggest that the promotion of the internalisation of the value of the task (such as publication) would be beneficial for research in astronomy. In order for that value to be internalised however, the way how publications are done may need to be adapted for them to better fit the quality requirements of astronomers (cf. Heuritsch, 2021b).

Future studies could inquire, together with astronomers, what such innovative publications could entail. Moreover, further research may uncover more aspects of research that would relate to astronomers' autonomous motivation and suggest how evaluation criteria could be designed on that basis. Building upon the results of that kind of research, could inspire incentives which leverage astronomers' autonomous motivation, and thereby increasing their well-being and work engagement. After all, "participation in planning and evaluation was [found to be] related to satisfaction, while participation in planning was related only to productivity" (Gagné et al., 2010).

**Supplementary Materials:** S1: Survey Questions, S2: EFAs and CFAs, S3: Descriptive Statistics

**Funding:** This study was performed in the framework of the junior research group "Reflexive Metrics", which is funded by the BMBF (German Bundesministerium für Bildung und Forschung; project number: 01PQ17002).

**Institutional Review Board Statement:** Not applicable.

**Informed Consent Statement:** Not applicable.

**Acknowledgments:** First: I would like to extend my gratitude to the 3509 astronomers who dedicated upwards of half an hour – despite the publish-or-perish imperative – to participate in this survey. Second, Thea Gronemeier and Florian Beng assisted the survey design and were a big support in data processing. Third, I would like to thank my supervisor, Stephan Gauch, for facilitating this project. Fourth, thank you to all the pre-testers: Niels Taubert, Jens Ambrasat, Andrej Dvornik, Iva Laginja, Levente Borvák, Nathalie Schwichtenberg, Theresa Velden, Richard Heidler, Rudolf Albrecht, Andreas Herdin, Alex Fenton and Philipp Löschl. Fifth, I would like to thank Marion Bräuer, whose graphic design expertise flowed into the visualisation of the motivation continuum and Nanou Haafs-Baart, who advised me on making the text more readable. Finally, I am grateful to GESIS (Lebniz Institut für Sozialwissenschaften) for the scholarship that enabled me to participate in their 2019 survey methodology course.

**Conflicts of Interest:** The authors declare no conflict of interest.

## References

Anderson, M.S.; Ronning, E.A.; De Vries, R.; Martinson, B.C. The Perverse Effects of Competition on Scientists' Work and Relationships. *Sci. Eng. Ethics* **2007**, *13*, 437–461. https://doi.org/10.1007/s11948-007-9042-5.

Bouter, L.M.; Tijdink, J.; Axelsen, N.; Martinson, B.C.; ter Riet, G. Ranking major and minor research misbehaviors: Results from a survey among participants of four World Conferences on Research Integrity. *Res. Integr. Peer Rev.* **2016**, *1*, 17.

Crain, L.A.; Martinson, B.C.; Thrush, C.R. Relationships between the Survey of Organizational Research Climate (SORC) and self-reported research practices. *Sci. Eng. Ethics* **2013**, *19*, 835–850.

Deci, E. L.; Ryan, R. M. Intrinsic motivation and self determination in human behaviour. New York, NY: Plenum **1985**.

Esser, H. Soziologie. Spezielle Grundlagen. Band 1: Situationslogik und Handeln. *KZfSS Kölner Z. Soziologie Soz.* **1999**, *53*, 773. https://doi.org/10.1007/s11577-001-0109-z.

Fochler, M.; De Rijcke, S. Implicated in the Indicator Game? An Experimental Debate. *Engag. Sci. Technol. Soc.* **2017**, *3*, 21–40.

Gagné, M.; Deci, E.L. Self-determination theory and work motivation. Journal of Organizational Behaviour, **2005**, 26, 331–362. doi:10.1002/job.322.

Gagné, M.; Forest, J.; Gilbert, M.-H.; Aubé, C.; Morin, E.; & Malorni, A. The Motivation at Work Scale: Validation evidence in two languages. Educational and Psychological Measurement, **2010**, 70, 628–646. doi:10.1177/0013164409355698.

Gagné, M.; Forest, J.; Vansteenkiste, M.; Crevier-Braud, L.; van den Broeck, A.; Aspeli, A.K.; Bellerose, J.; Benabou, C.; Chemolli, E.; Güntert, S.T.; Halvari, H.; Laksmi Indiyastuti, D.; Johnson, P.A.; Hauan Molstad, M.; Naudin, M.; Ndao, A.; Hagen Olafsen, A.; Roussel, P.; Wang, Z.; Westbye, C. The Multidimensional Work Motivation Scale: Validation evidence in seven languages and nine countries, European Journal of Work and Organizational Psychology, **2015**, 24:2, 178-196, DOI: 10.1080/1359432X.2013.877892.
Halffman, W.; Radder, H. The Academic Manifesto: From an Occupied to a Public University. *Minerva* **2015**, *53*, 165–187.

Haven, T.L. Towards a Responsible Research Climate: Findings from Academic Research in Amsterdam. 2021. Available online: https://research.vu.nl/en/publications/towards-a-responsible-research-climate-findings-from-academic-res (accessed on 26th November 2021).

Heuritsch, J. Effects of metrics in research evaluation on knowledge production in astronomy A case study on Evaluation Gap and Constitutive Effects. In Proceedings of the STS Conference Graz 2019, Graz, Austria, 6–7 May 2019; https://doi.org/10.3217/978-3-85125-668-0-09.

Heuritsch, J. Reflexive Behaviour: How Publication Pressure Affects Research Quality in Astronomy. *Publications* **2021a**, *9*, 52. https://doi.org/10.3390/publications9040052.

Heuritsch, J. The Evaluation Gap in Astronomy—Explained through a Rational Choice Framework. *arXiv* **2021b**, arXiv:2101.03068.

Martinson, B.C.; Anderson, M.S.; Crain, A.L.; De Vries, R. Scientists' perceptions of organizational justice and self-reported misbehaviors. *J. Empir. Res. Hum. Res. Ethics* **2006**, *1*, 51–66.

Martinson, B.C.; Anderson, M.S.; de Vries, R. Scientists behaving badly. *Nature* **2005**, *435*, 737–738.

Martinson, B.C.; Crain, A.L.; Anderson, M.S.; De Vries, R. Institutions 'Expectations for Researchers' Self-Funding, Federal Grant Holding, and Private Industry Involvement: Manifold Drivers of Self-Interest and Researcher Behavior. *Acad. Med.* **2009**, *84*, 1491–1499.

Martinson, B.C.; Crain, L.A.; De Vries, R.; Anderson, M.S. The importance of organizational justice in ensuring research integrity. *J. Empir. Res. Hum. Res. Ethics* **2010**, *5*, 67–83.

Ryan, R. M.; Connell, J. P. Perceived locus of causality and internalization: Examining reasons for acting in two domains. Journal of Personality and Social Psychology, **1989**, 57, 749–761. doi:10.1037/0022-3514.57.5.749.

Siegrist, J.; Li, J.; Montano, D. Psychometric Properties of the Effort-Reward Lmbalance Questionnaire; Department of Medical Sociology, Faculty of Medicine, Duesseldorf University: Düsseldorf, Germany, 2014.

Tijdink, J.K.; Smulders, Y.M.; Vergouwen, A.C.M.; de Vet, H.C.W.; Knol, D.L. The assessment of publication pressure in medical science; validity and reliability of a Publication Pressure Questionnaire (PPQ). *Qual. Life Res.* **2014**, *23*, 2055–2062.

# S1: Survey Questions

**S1-TableS1: Item-battery for the IV M1: "Motivation to become an Astronomer".**

Question:

"How important were the following motivational factors for you to become an astronomer?"

Items:

Items were designed by the author, and are inspired by Gagné et al. (2015).
For each question the response scale ranged from 1=Strongly Disagree to 5=Strongly Agree.

| |
|---|
| Out of curiosity |
| I needed a job |
| My goal is to find out more about the laws that govern the universe |
| I enjoy the process of gaining insight in astronomical phenomena |
| Astronomy is a prestigious field in science |
| Being a scientist is a prestigious job |
| I like the intellectual challenge |
| I find basic research more gratifying than the sometimes more profit-oriented activities in other natural sciences |

**S1-TableS2a: Item-battery for the IV M2: "Drivers to Publish".**

Question:

“What are your personal drivers to publish papers?”

Items:

Items were designed by the author, and are inspired by Gagné et al. (2015).
For each question the response scale ranged from 1=Strongly Disagree to 5=Strongly Agree.

| Publishing is important to share results with the community |
| --- |
| I feel ashamed if I don’t publish |
| Publishing is a requirement from my job |
| Publishing enhances my career prospects |
| I enjoy the review process |
| Publishing my results makes me proud of myself |
| Writing results down has personal significance to me |
| Publishing increases my reputation as a scientist |
| I enjoy the process of writing a paper |

**S1-TableS2b: Item-battery for the IV M3: “Feelings one experiences when not publishing the amount of papers that one aimed to publish”.**

Question:

“How do you feel when you don’t publish the amount of papers that you aimed to publish?”

Items:

Items were designed by the author, and are inspired by Gagné et al. (2015).
For each question the response scale ranged from 1=Strongly Disagree to 5=Strongly Agree and included the option “NA”.

| I feel ashamed |
| --- |
| I feel like I am not a good researcher |
| I feel like I am not doing a good job |
| I feel worthless |
| I am worried that it will negatively impact my career prospects |
| That’s the risk of research that sometimes you are stuck, so I don’t feel any negative emotions |
| I feel disappointed that I cannot share any new insights with my community |
| I am worried that it will negatively impact my research track record |
| I am worried that it will decrease my chances for receiving external grants |
| I am worried that it will decrease my chances for receiving telescope time |

**S1-TableS3: Item-battery for the instrument “Source of the Perceived Publication Pressure”.**

Question:

“What is the source of that pressure? I feel publication pressure, because …”

Items:

Items were designed by the author and were only asked if the survey respondent answered more than “never” in a previous question about “How often do you feel pressure to publish?” (answer options: “never”, “very rarely”, “rarely”, “regularly”, “often”, “very often”).
This was a multiple choice question, where each item could be selected or not selected.

| ... I need to meet my supervisor’s/ boss’s expectations |
| --- |
| ... I need to boost my publication record for increasing my career chance |

| ... publishing gives me a sense of accomplishment |
| --- |
| ... I need to earn prestige |
| ... I need to avoid failure |
| ... of the need to publish first |
| ... only getting results out will push knowledge forward |
| ... I need to maintain credibility as a scientist |
| ... of the policies of my institute |

**S1-TableS4: Item-battery for the instrument "Most rewarding aspects about work".**

Question:

"What do you find most rewarding about your work?"

Items:

Items were designed by the author.
This was a ranking question, where a survey respondent chose 3 answers in a ranked order.

| Enjoying the process of finding truths about the universe |
| --- |
| Receiving praise from a colleague/ my supervisor |
| Making incremental steps in building up knowledge |
| Making ground-breaking steps in building up knowledge |
| Winning scientific prizes |
| Receiving a job promotion (a more senior job title) |

| Receiving a salary raise |
| --- |
| Getting a paper published |

## S2: EFAs & CFAs

### S2-TableS1a: M1 – CATPCA (4 factors)

This table displays the SPSS output of the 4-factor CATPCA of the M1 construct consisting of 8 items (see *Table S1 in S1*).

**Pattern Matrix**

| M1 Items | Dimension | | | |
|---|---|---|---|---|
| | Introjected Regulation | Identified Regulation | Intrinsic Motivation | External Regulation |
| Out of curiosity | -.203 | -.025 | .686 | .451 |
| I needed a job | .203 | .015 | -.234 | .862 |
| My goal is to find out more about the laws that govern the universe | .085 | .076 | .667 | -.153 |
| I enjoy the process of gaining insight in astronomical phenomena | -.011 | .931 | .010 | .002 |
| Astronomy is a prestigious field in science | .908 | .012 | .070 | .088 |
| Being a scientist is a prestigious job | .895 | -.020 | .088 | .113 |
| I like the intellectual challenge | .003 | .926 | .020 | .014 |
| I find basic research more gratifying than the sometimes more profit-oriented activities in other natural sciences | .209 | -.037 | .688 | -.265 |

### S2-TableS1b: M1 – Types of Motivation; Theory versus CATPCA

Comparison between the categorisation as expected from theory and as obtained from the CATPCA for the types of motivation of the M1 construct. Differing results are denoted by a *.

| M1 Items | Type of Motivation (as expected from theory) | Type of Motivation (results from CATPCA; *from TableS1a in S2*) |
|---|---|---|
| Out of curiosity | Intrinsic Motivation | Intrinsic Motivation |
| I needed a job | External Regulation | External Regulation |
| My goal is to find out more about the laws that govern the universe | Identified Regulation | Intrinsic Motivation* |
| I enjoy the process of gaining insight in astronomical phenomena | Intrinsic Motivation | Identified Regulation* |
| Astronomy is a prestigious field in science | Introjected Regulation | Introjected Regulation |
| Being a scientist is a prestigious job | Introjected Regulation | Introjected Regulation |
| I like the intellectual challenge | Intrinsic Motivation | Identified Regulation* |
| I find basic research more gratifying than the sometimes more profit-oriented activities in other natural sciences | Identified Regulation | Intrinsic Motivation* |

**S2-TableS1c: M1 – Cronbach Alphas before removal of items**

This table displays the SPSS output of the Cronbach alphas of the M1 construct consisting of 8 items. * denotes items that we subsequently removed.

**Reliability Statistics**

| Cronbachs Alpha | N of Items |
|---|---|
| .583 | 8 |

**Item-Total-Statistics**

| M1 Items | Scale Mean if Item Deleted | Scale Variance if Item Deleted | Corrected Item-Total Correlation | Cronbach's Alpha if Item Deleted |
|---|---|---|---|---|
| Out of curiosity* | 25.56 | 16.360 | .187 | .582 |
| I needed a job* | 27.35 | 16.976 | .060 | .633 |
| My goal is to find out more about the laws that govern the universe | 25.48 | 16.022 | .290 | .550 |
| I enjoy the process of gaining insight in astronomical phenomena | 25.10 | 16.530 | .358 | .540 |
| Astronomy is a prestigious field in science | 26.53 | 13.685 | .430 | .497 |
| Being a scientist is a prestigious job | 26.66 | 13.743 | .442 | .493 |
| I like the intellectual challenge | 25.16 | 16.416 | .372 | .536 |
| I find basic research more gratifying than the sometimes more profit-oriented activities in other natural sciences | 25.68 | 15.607 | .283 | .552 |

**S2-TableS1d: cleaned M1 – Cronbach Alphas after removal of items**

This table displays the SPSS output of the Cronbach alphas of the cleaned M1 construct after removal of items, which resulted in 6 remaining items.

**Reliability Statistics**

| Cronbachs Alpha | N of Items |
|---|---|
| .648 | 6 |

**Item-Total-Statistics**

| Cleaned M1 Items | Scale Mean if Item Deleted | Scale Variance if Item Deleted | Corrected Item-Total Correlation | Cronbach's Alpha if Item Deleted |
|---|---|---|---|---|
| My goal is to find out more about the laws that govern the universe | 19.10 | 11.049 | .317 | .626 |
| I enjoy the process of gaining insight in astronomical phenomena | 18.72 | 11.546 | .390 | .608 |
| Astronomy is a prestigious field in science | 20.15 | 9.055 | .455 | .573 |
| Being a scientist is a prestigious job | 20.29 | 9.211 | .452 | .575 |
| I like the intellectual challenge | 18.79 | 11.570 | .378 | .611 |
| I find basic research more gratifying than the sometimes more profit-oriented activities in other natural sciences? | 19.30 | 10.589 | .321 | .627 |

**S2-TableS1e: cleaned M1 – CATPCA (2 factors)**

This table displays the SPSS output of the 2-factor CATPCA of the cleaned M1 construct after removal of items, which resulted in 6 remaining items.

**Pattern Matrix**

| | **Dimension** | |
|---|---|---|
| Cleaned M1 Items | M1F1- Autonomous Motivation | M1F2 - Controlled Motivation |
| My goal is to find out more about the laws that govern the universe | .610 | .030 |
| I enjoy the process of gaining insight in astronomical phenomena | .882 | -.059 |
| Astronomy is a prestigious field in science | .028 | .919 |
| Being a scientist is a prestigious job | .003 | .919 |
| I like the intellectual challenge | .867 | -.045 |
| I find basic research more gratifying than the sometimes more profit-oriented activities in other natural sciences? | .469 | .184 |

**S2-TableS2a: M2 – CATPCA (4 factors)**

This table displays the SPSS output of the 4-factor CATPCA of the M2 construct consisting of 9 items (see *Table S2a in S1*).

**Pattern Matrix**

| | **Dimension** | | | |
|---|---|---|---|---|
| M2 Items | Intrinsic Motivation | External Regulation | Introjected Regulation | Identified Regulation |
| Publishing is important to share results with the community | -.104 | .101 | -.123 | .978 |
| I feel ashamed if I don't publish | -.185 | .010 | .798 | -.302 |
| Publishing is a requirement from my job | .001 | .911 | -.170 | .104 |
| Publishing enhances my career prospects | .051 | .722 | .255 | -.023 |
| I enjoy the review process | .961 | .152 | -.105 | -.293 |
| Publishing my results makes me proud of myself | .129 | -.114 | .739 | .133 |
| Writing results down has personal significance to me | .492 | -.139 | .221 | .301 |
| Publishing increases my reputation as a scientist | .003 | .384 | .512 | .128 |
| I enjoy the process of writing a paper | .807 | -.062 | -.053 | .108 |

**S2-TableS2b: M2 – Types of Motivation; Theory versus CATPCA**

Comparison between the categorisation as expected from theory and as obtained from the CATPCA for the types of motivation of the M2 construct. Differing results are denoted by a *.

| **M2 Items** | **Type of Motivation (as expected from theory)** | **Type of Motivation (results from CATPCA;** *from TableS2a in S2***)** |
|---|---|---|
| Publishing is important to share results with the community | Identified Regulation | Identified Regulation |
| I feel ashamed if I don't publish | Introjected Regulation | Introjected Regulation |
| Publishing is a requirement from my job | External Regulation | External Regulation |
| Publishing enhances my career prospects | External Regulation | External Regulation |
| I enjoy the review process | Intrinsic Motivation | Intrinsic Motivation |
| Publishing my results makes me proud of myself | Introjected Regulation | Introjected Regulation |
| Writing results down has personal significance to me | Identified Regulation | Intrinsic Motivation* |
| Publishing increases my reputation as a scientist | Introjected Regulation | Introjected Regulation |
| I enjoy the process of writing a paper | Intrinsic Motivation | Intrinsic Motivation |

**S2-TableS2c: M2 – Cronbach Alphas**

This table displays the SPSS output of the Cronbach alphas of the M2 construct consisting of 9 items.

**Reliability Statistics**

| Cronbachs Alpha | N of Items |
|---|---|
| .652 | 9 |

**Item-Total-Statistics**

| M2 Items | Scale Mean if Item Deleted | Scale Variance if Item Deleted | Corrected Item-Total Correlation | Cronbach's Alpha if Item Deleted |
|---|---|---|---|---|
| Publishing is important to share results with the community | 28.05 | 22.240 | .265 | .639 |
| I feel ashamed if I don't publish | 29.40 | 20.768 | .198 | .661 |
| Publishing is a requirement from my job | 28.59 | 21.078 | .232 | .647 |
| Publishing enhances my career prospects | 28.62 | 19.886 | .384 | .612 |
| I enjoy the review process | 30.13 | 20.598 | .297 | .632 |
| Publishing my results makes me proud of myself | 28.75 | 19.082 | .477 | .590 |
| Writing results down has personal significance to me | 29.03 | 19.582 | .396 | .609 |
| Publishing increases my reputation as a scientist | 28.59 | 19.745 | .459 | .598 |
| I enjoy the process of writing a paper | 29.35 | 20.024 | .310 | .630 |

**S2-TableS2d: M2 – CATPCA (2 factors)**

This table displays the SPSS output of the 2-factor CATPCA of the M2 construct consisting of 9 items.

**Pattern Matrix**

| M2 Items | Dimension | |
|---|---|---|
| | M2F1- Autonomous Motivation | M2F2 - Controlled Motivation |
| Publishing is important to share results with the community | .544 | -.008 |
| I feel ashamed if I don't publish | -.122 | .591 |
| Publishing is a requirement from my job | -.176 | .683 |
| Publishing enhances my career prospects | -.030 | .811 |
| I enjoy the review process | .670 | -.067 |
| Publishing my results makes me proud of myself | .488 | .424 |
| Writing results down has personal significance to me | .752 | .011 |
| Publishing increases my reputation as a scientist | .174 | .714 |
| I enjoy the process of writing a paper | .823 | -.169 |

**S2-TableS3a: M3 – CATPCA (3 factors)**

This table displays the SPSS output of the 3-factor CATPCA of the M3 construct consisting of 10 items (see *Table S2b in S1*).

**Pattern Matrix**

| M3 Items | Dimension | | |
|---|---|---|---|
| | Introjected Regulation | External Regulation | Residual Category |
| I feel ashamed | .891 | -.042 | .074 |
| I feel like I am not a good researcher | .886 | -.007 | .055 |
| I feel like I am not doing a good job | .851 | -.005 | .138 |
| I feel worthless | .885 | -.048 | .074 |
| I am worried that it will negatively impact my career prospects | .138 | .802 | -.103 |
| That's the risk of research that sometimes you are stuck, so I don't feel any negative emotions | -.589 | -.134 | .380 |
| I feel disappointed that I cannot share any new insights with my community | .158 | .069 | .939 |
| I am worried that it will negatively impact my research track record | .138 | .811 | -.088 |
| I am worried that it will decrease my chances for receiving external grants | -.107 | .907 | .046 |
| I am worried that it will decrease my chances for receiving telescope time | -.162 | .708 | .307 |

**S2-TableS3b: M3 – Types of Motivation; Theory versus CATPCA**

Comparison between the categorisation as expected from theory and as obtained from the CATPCA for the types of motivation of the M3 construct.

There is no difference between the categorisation based on theory and the factors of the CATPCA (see *TableS3a in S2*). The residual category of the CATPCA consists of the single item representing identified regulation and the control item (in italics). Pearson correlations of all items with the control item were as expected negative and this item was subsequently removed from further analysis.

| M3 Items | Type of Motivation |
|---|---|
| I feel ashamed | Introjected Regulation |
| I feel like I am not a good researcher | Introjected Regulation |
| I feel like I am not doing a good job | Introjected Regulation |
| I feel worthless | Introjected Regulation |
| I am worried that it will negatively impact my career prospects | External Regulation |
| That's the risk of research that sometimes you are stuck, so I don't feel any negative emotions | *Control Item* |
| I feel disappointed that I cannot share any new insights with my community | Identified Regulation |
| I am worried that it will negatively impact my research track record | External Regulation |
| I am worried that it will decrease my chances for receiving external grants | External Regulation |
| I am worried that it will decrease my chances for receiving telescope time | External Regulation |

**S2-TableS3c: cleaned M3 – Cronbach Alphas**

This table displays the SPSS output of the Cronbach alphas of the cleaned M3 construct consisting of 9 items.

**Reliability Statistics**

| Cronbachs Alpha | N of Items |
|---|---|
| .871 | 9 |

**Item-Total-Statistics**

| Cleaned M3 Items | Scale Mean if Item Deleted | Scale Variance if Item Deleted | Corrected Item-Total Correlation | Cronbach's Alpha if Item Deleted |
|---|---|---|---|---|
| I feel ashamed | 26.07 | 51.715 | .700 | .848 |
| I feel like I am not a good researcher | 25.67 | 51.424 | .721 | .846 |
| I feel like I am not doing a good job | 25.44 | 52.724 | .694 | .849 |
| I feel worthless | 26.29 | 52.784 | .659 | .852 |
| I am worried that it will negatively impact my career prospects | 25.12 | 52.826 | .678 | .851 |
| I feel disappointed that I cannot share any new insights with my community | 25.51 | 60.606 | .318 | .880 |
| I am worried that it will negatively impact my research track record | 25.07 | 53.924 | .679 | .851 |
| I am worried that it will decrease my chances for receiving external grants | 24.80 | 56.044 | .567 | .861 |
| I am worried that it will decrease my chances for receiving telescope time | 25.44 | 57.699 | .438 | .872 |

**S2-TableS3d: cleaned M3 – CATPCA (3 factors)**

This table displays the SPSS output of the 3-factor CATPCA of the cleaned M3 construct consisting of 9 items.

**Pattern Matrix**

| Cleaned M3 Items | **Dimension** | | |
|---|---|---|---|
| | M3F2- Introjected Regulation | M3F3- External Regulation | M3F1- Identified Regulation |
| I feel ashamed | .899 | -.021 | .001 |
| I feel like I am not a good researcher | .896 | .015 | -.009 |
| I feel like I am not doing a good job | .852 | .013 | .085 |
| I feel worthless | .896 | -.028 | .011 |
| I am worried that it will negatively impact my career prospects | .147 | .812 | -.126 |
| I feel disappointed that I cannot share any new insights with my community | .067 | .020 | 1.032 |
| I am worried that it will negatively impact my research track record | .139 | .817 | -.082 |
| I am worried that it will decrease my chances for receiving external grants | -.095 | .903 | .012 |
| I am worried that it will decrease my chances for receiving telescope time | -.158 | .690 | .267 |

**S2-TableS4: Results of the comparative factor analyses (CFAs) our motivation constructs M1-M3.**

Note: These results take significant covariations between indicators into account.

| **Independent Variable** | **Comparative Fit Index (CFI)** | **Tucker-Lewis Index (TLI)** | **Root Mean Square Error of Approximation (RMSEA)** |
|---|---|---|---|
| M1 | 0.994 | 0.988 | 0.032 90%CI(.018, .046). |
| M2 | 0.945 | 0.891 | 0.071 90%CI(.062, .079). |
| M3 | 0.982 | 0.972 | 0.060 90%CI(.050, .070). |

# S3: Descriptive Statistics

### S3-TableS1: Descriptive Statistics of M1.

This table displays the SPSS output of the descriptive statistics of the M1 construct consisting of 8 items (see *Table S1 in S1*) and its factors (see *Table S1e in S2*).

| Factor | Item | N | Mean | Sdt.-Deviation |
|---|---|---|---|---|
| *Removed* | Out of curiosity | 2496 | 4.09 | 1.124 |
| *Removed* | I needed a job | 2494 | 2.30 | 1.303 |
| M1F1 | My goal is to find out more about the laws that govern the universe | 2501 | 4.17 | .992 |
| M1F1 | I enjoy the process of gaining insight in astronomical phenomena | 2507 | 4.55 | .757 |
| M1F2 | Astronomy is a prestigious field in science | 2502 | 3.12 | 1.268 |
| M1F2 | Being a scientist is a prestigious job | 2496 | 2.98 | 1.236 |
| M1F1 | I like the intellectual challenge | 2502 | 4.48 | .765 |
| M1F1 | I find basic research more gratifying than the sometimes more profit-oriented activities in other natural sciences | 2501 | 3.97 | 1.109 |

### S3-TableS2: Descriptive Statistics of M2.

This table displays the SPSS output of the descriptive statistics of the M2 construct consisting of 9 items (see *Table S2a in S1*) and its factors (see *Table S2d in S2*).

| Factor | Item | N | Mean | Sdt.-Deviation |
|---|---|---|---|---|
| M2F1 | Publishing is important to share results with the community | 2035 | 4.51 | .746 |
| M2F2 | I feel ashamed if I don't publish | 2030 | 3.16 | 1.269 |
| M2F2 | Publishing is a requirement from my job | 2034 | 3.97 | 1.109 |
| M2F2 | Publishing enhances my career prospects | 2032 | 3.94 | 1.065 |
| M2F1 | I enjoy the review process | 2032 | 2.44 | 1.081 |
| M2F1 | Publishing my results makes me proud of myself | 2033 | 3.81 | 1.069 |
| M2F1 | Writing results down has personal significance to me | 2030 | 3.54 | 1.103 |
| M2F2 | Publishing increases my reputation as a scientist | 2034 | 3.97 | .974 |
| M2F1 | I enjoy the process of writing a paper | 2034 | 3.22 | 1.175 |

**S3-TableS3: Descriptive Statistics of M3.**

This table displays the SPSS output of the descriptive statistics of the M3 construct consisting of 10 items (see *Table S2b in S1*) and its factors (see *Table S3d in S2*).

| Factor | Item | N | Mean | Sdt.-Deviation |
|---|---|---|---|---|
| M3F2 | I feel ashamed | 1893 | 2.59 | 1.397 |
| M3F2 | I feel like I am not a good researcher | 1921 | 3.01 | 1.388 |
| M3F2 | I feel like I am not doing a good job | 1924 | 3.26 | 1.306 |
| M3F2 | I feel worthless | 1884 | 2.38 | 1.368 |
| M3F3 | I am worried that it will negatively impact my career prospects | 1890 | 3.63 | 1.308 |
| *removed* | That's the risk of research that sometimes you are stuck, so I don't feel any negative emotions | 1881 | 2.78 | 1.203 |
| M3F1 | I feel disappointed that I cannot share any new insights with my community | 1919 | 3.19 | 1.168 |
| M3F3 | I am worried that it will negatively impact my research track record | 1910 | 3.63 | 1.223 |
| M3F3 | I am worried that it will decrease my chances for receiving external grants | 1836 | 3.91 | 1.187 |
| M3F3 | I am worried that it will decrease my chances for receiving telescope time | 1510 | 3.26 | 1.271 |